\documentstyle[12pt]{article}
\begin{document}

\author{E. N. Antonov  \\
{\em Petersburg Nuclear Physics Institute,}\\ {\em Gatchina, 188350, Russia.}}
\title{Beyond Low-Gribov theorem for high energy interactions of scalar
and gauge particles.\thanks{
Work supported by the Russian Fund of Fundamental Research, grant
93-02-16809.}}
\maketitle

\begin{abstract}
We obtain a generalization of the Low theorem for non-Abelian boson emission
in collision of scalar and gauge vector particles and its extension to high
energy collisions for small transverse momenta of produced particles. We
demonstrate that in the case of particles with spin the direct extension the
Low formula to high energy is in contradiction with the correct amplitude
factorization behavior. Consideration of different kinematical regions and
use of methods of dual models allows us to separate
contributions of intermediate excited states and standard spin
corrections in the Low formulae. We show that the amplitude
factorization occurs at high energy due to the contribution
of the intermediate states which is additional to the gluon
production amplitude for the scalar particle collision.
\end{abstract}

\newpage

\section{Introduction.}

In recent years there has been significant interest to the experimental and
theoretical studies of deep inelastic scattering \cite{1} when provided with
very useful information concerning the further development of high-energy
physics. In turn, the Low theorem \cite{2} gives us low energy static
characteristics like charges, magnetic moments and others. As well known,
one can extend its applicability to the region of high-energy collisions
\cite{3}. In QED, it proves possible to calculate the inelastic amplitude
for the photon production with a small transverse momentum in terms of the
elastic amplitude and its derivatives. The next steps were to consider the
non-Abelian generalization and to take into account the contribution of
excited intermediate states in the corresponding channels (in
the channels $ s_1,s_{1^{\prime }},s_2,$ and $s_{2^{\prime }}$
in Fig.1 where $ s_1=-2(p_1k),\;s_{1^{\prime }}=2\left(
p_4k\right) ,\;s_2=-2(p_2k)$ and $ s_{2^{\prime }}=2\left(
p_3k\right) $ ). These problems have been considered by Lipatov
for scalar case both in the multi-Regge kinematics \cite{4} ($
s\gg s_i\gg m_{ch}^2$, where a $m_{ch}$ is the characteristic mass scale of
elementary elastic process in Fig.1) and in the fragmentation region \cite{5}
$\left( s\gg s_2\sim s_{2^{\prime }}\sim m_{ch}^2\right) $. The gauge
massless particles (gluons or gravitons) are radiated in collisions of
scalar particles. There the inelastic amplitude can be considered as the
dispersion representation in the variables $s_1,s_{1^{\prime }},s_2,$ and $
s_{2^{\prime }}$. The Low formula can be considered as the contribution of
the ground scalar states in this dispersion relation as pole in $s_i=0$. The
contributions of the excited intermediate states is represented by the
dispersion integral with subtraction in $s_i=0$. Such the representation
allows the low and high energy regions to be consistent . At low energy the
contribution of intermediate states dies out and we have the Low formula.
At high energy the contribution of the intermediate states is necessary for
the correct analytical amplitude behavior. The general condition of
analyticity of the inelastic amplitude in the $s_1,s_{1^{\prime }},s_2$ and
$s_{2^{\prime }}$ fixes the contribution of excited intermediate states up
to an additive constant which provide the subtraction of the dispersion
integral in $s_i=0$ . The constant depends on a chosen model and may depend
on a sort of external on-shell particles in the given channel. The value of
the constant in Ref. \cite{4,5} was estimated in terms of the open and
closed string models. It equals $\Gamma ^{\prime }\left( 1\right) =-c_E$
for the open string and $2\log 2-c_E$ for the closed one where $c_E$ is the
Euler constant. In this case the gluon or the graviton plays a role of the
gauge particle and tachyons correspond to on-shell scalar particles. The
next step was to consider scattering of gauge particles where the gluon and
the graviton were considered as the next excited states for open and closed
strings. Such an inelastic amplitude for five gauge massless particles has
been calculated in Ref. \cite{6} in the multi-Regge kinematics. At first
sight, it seems that the additive constant is the same as for the tachyon
collision because of the amplitude factorization, but in fact there exist
some additional terms which might contribute to it. In this kinematics we do
not have any obvious idea how to single out the contribution of excited
intermediate states to clarify the situation. It is necessary to consider
another kinematical region, a fragmentation region.

The reason to deal with the amplitude including gauge and scalar particles
is to compare contributions of the bremsstrahlung radiation from scalar and
vector particle and to find the dependence of the additive constant on the
nature of on-shell external particles. Moreover, we would like to
investigate the transformation of every contribution in detail, using the
string case as a guide, and going from small $s_i$ (the Low formulae are
valid) through the fragmentation region $s_i\sim m_{ch}^2\sim \frac 1{\alpha
^{\prime }}$ (where $\alpha ^{\prime }$ is a slope of Regge trajectory and
the contribution of intermediate particles is yet to be controlled) to large
$s_i\gg m_{ch}^2\sim \frac 1{\alpha ^{\prime }}$ (multi-Regge kinematics).

This paper is also concerned with the contribution of gauge non-Abelian
particles to the Low theorem in the field theory. We demonstrate that in the
case of particles with spin the direct extension the Low formula to high
energy is in contradiction with the correct amplitude factorization
behavior. Comparison of its continuation to the Regge kinematics (where the
perturbative theory is valid) with string expressions provides with an
additional test (in particular, in the sense of limit $\alpha ^{\prime
}\rightarrow 0$ \cite{12}) and shows that it is necessary to take into
account the contribution of intermediate particles additional to the scalar
scattering case. It is the use of the bosonic string theory which allows us
to control all contributions in different kinematical regions. In doing so,
in contrast with loop calculations \cite{13}, Born approximation appears to
be enough.

Section (2) deals with the Low theorem for scalar and vector Yang-Mills
particles. We extend it to different kinematics in agreement with Gribov's
requirement and demonstrate that some corrections are in contradiction with
correct amplitude factorization. In section (3) we consider the same process
in the string theory in the multi-Regge kinematics where we take into
account the excited intermediate states and obtain the amplitude with the
correct factorization. The fragmentation region is considered in section
(4), which allows us to separate the contribution of the all intermediate
excited states and ground states and to consider the all kinematical
regions. As a result we observed that the restoration of the amplitude
factorization is due to a new additional type of the intermediate excited
states.

\section{Low theorem in the case of Yang-Mills scalar and vector particles.}

The Low theorem, owing to gauge invariance, allows us to obtain a soft
vector particle production amplitude in terms of the elastic one and its
derivatives. Our presentation follows \cite{7,8}. Now we are going to
generalize this procedure for a non-Abelian field theory. In this case the
Lagrangian is given by
\begin{equation}
\label{4}
\begin{array}{c}
L=-\frac 14\left( \partial _\mu A_\nu ^a-\partial _\nu A_\mu
^a-gf_{bc}^aA_\mu ^bA_\nu ^c\right) ^2\qquad \qquad \qquad \qquad \\
\\
+\left[ \left( \partial _\mu -gA_{\mu \ a}T^a\right) \Phi \right] ^{+}\left[
\left( \partial ^\mu -gA_a^\mu T^a\right) \Phi \right] +\cdots \ ,
\end{array}
\end{equation}

where $T^a$ are generators of $SU\left( N\right) $ gauge group in the
adjoint representation in which their matrix elements are given as structure
constants of this group
\begin{equation}
\label{5}
\begin{array}{c}
\left( T^a\right) _{bc}=f_{bc}^a=f_c^{ab}=f_{abc} \\
\\
\left[ T^a,T^b\right] =if_c^{ab}T^c
\end{array}
{}.
\end{equation}

The basic elastic amplitude with one vector and three scalar particles (see
Fig.3) will be denoted $l_{2\ \nu }A^{\nu ,\ a_1,a_2,a_3,a_4}\left(
p_1,p_2,p_3,p_4\right) $ where $l_{2\ \nu }$ is the polarization vector of
the gauge particle, $a_i$ are color indices of the gauge group and $p_i$ are
the corresponding momenta of particles. The inelastic amplitude is
represented by (see Fig.4)
$$
A_g\left( l,k,\ldots \right) =A_g^{ext}\left( l,k,\ldots \right)
+A_g^{int}\left( k,l,\ldots \right) \ .
$$
where $l$ and $k$ are the polarization vector and the momentum of the
radiated gluon with the color index $a$ respectively. $A_g^{int}$ is the
contribution of diagrams in which an gluon is radiated from an internal
line. The sum of pole diagrams such that the additional gluon is radiated
from an external line is given in the form
\begin{equation}
\label{6}
\begin{array}{c}
A_g^{ext\ a,\ a_1,a_2,a_3,a_4}\left( l,k;p_1;l_2,p_2;p_3;p_4\right) = \\
\\
=ig\sum_cf_c^{a\;a_2}
\frac{\left( l_{2\ \nu }\left( lp_2\right) -l_{\ \nu }\left( l_2k\right)
+\left( l_2l\right) k_{\ \nu }\right) }{kp_2}A^{\nu ,\ a_1,c,a,a_5}\left(
\ldots ,p_2+k,\ldots \right) + \\  \\
+ig\sum_{i\neq 2}\sum_{c_i}f_{c_i}^{a\;a_i}\frac{\left( lp_i\right) }{kp_i}
\left( l_{2\ \nu }A^{\nu ,\ \ldots ,c_{i,\ldots }}\left( \ldots
,p_i+k,\ldots \right) \right) \qquad \qquad \qquad
\end{array}
\end{equation}

All particles are incoming, so that $\sum_ip_i=-k$. Using the gauge
invariance $l\rightarrow l+ck$ we note that
\begin{equation}
\label{7}
\begin{array}{c}
A_g^{ext\ a,\ a_1,a_2,a_3,,a_4}\left( l=k,k;\ldots ;\ldots \right) = \\
\\
=ig\sum_i\sum_{c_i}f_{c_i}^{a\;a_i}\left( l_{2\ \nu }A^{\nu ,\ \ldots
,c_i,\ldots }\left( \ldots ,p_i+k,\ldots \right) \right)
\end{array}
\end{equation}

Following Ref.\cite{7}, we now define more carefully a procedure for the
extrapolation from elastic $A^{\nu \ \ldots ,c_i,\ldots }\left( \ldots
,p_i^{\prime },\ldots \right) $ with a physically realizable set of momenta $
\{p_i^{\prime }\}$ such that $\sum_ip_i^{\prime }=0$ and $p_i^{\prime
\;2}=m_i^2$ to $A^{\nu \ \ldots ,c_{i,\ldots }}\left( \ldots ,p_i,\ldots
\right) $ of the form given in Eq.(\ref{6}), where one momentum is
unphysical and the rest are the same as in $A_g$. Let $p_i^{\prime }=p_i-\xi
_i\left( k\right) $, then the vectors $\xi _i$ must have the properties $
\sum_i\xi _i\left( k\right) =-k,\left( p_i,\xi \right) =0,\xi _i\left(
0\right) =0,\left( \frac{\partial \xi _i}{\partial k_{\ \mu }}\right) _{k_{\
\mu }=0}=c<\infty .$ It is possible for the $\xi $'s to be defined so that
scalar variables are the same to first order in $k$ whether expressed in
terms of $p_i$ or $p_i^{\prime }$.

Now we can derive the relation between the elastic amplitude with
on-mass-shell momenta and the elastic amplitude with one unphysical momentum
\begin{equation}
\label{8}
\begin{array}{c}
A^{\nu ,\ \ldots ,c_{i,\ldots }}\left( \ldots ,p_i+k_{,\ldots }\right)
=A^{\nu ,\ \ldots ,c_{i,\ldots }}\left( \ldots ,p_i^{\prime }{}_{,\ldots
}\right) + \\
\\
\sum_j\xi _j\frac \partial {\partial p_j}A^{\nu ,\ \ldots ,c_{i,\ldots
}}\left( \ldots ,p_i{}_{,\ldots }\right) \mid _{\left\{ p_i\right\} =\left\{
p_i^{\prime }\right\} }+k\frac \partial {\partial p_i}A^{\nu ,\ \ldots
,c_{i,\ldots }}\left( \ldots ,p_i{}_{,\ldots }\right) \mid _{_{\left\{
p_i\right\} =\left\{ p_i^{\prime }\right\} }}.
\end{array}
\end{equation}

As a result of charge conservation we have
\begin{equation}
\label{9}
\begin{array}{c}
A_g^{ext\ a,\ a_1,a_2,a_3,a_4}\left( l=k,\;k;\ldots \right) = \\
\\
=ig\sum_i\sum_{c_i}f_{c_i}^{a\;a_i}\left( l_{2\ \nu }\left( k\frac \partial
{\partial p_i}\right) A^{\nu ,\ \ldots ,c_i,\ldots }\left( \ldots
,p_i,\ldots \right) \mid _{\left\{ p_i\right\} =\left\{ p_i^{\prime
}\right\} }\right) .
\end{array}
\end{equation}

Consequently, $k$ - independent terms $O((k)^0)$ which may come from either $
A_g^{ext\ }$ or $A_g^{int}$ are completely determined by the
gauge-invariance requirement (for QED see \cite{8}) $A_g\left( l=k,k,\ldots
\right) =0$ is given by
\begin{equation}
\label{10}-ig\sum_i\sum_{c_i}f_{c_i}^{a\;a_i}l_{2\ \nu }\left( l\frac
\partial {\partial p_i}\right) A^{\nu ,\ \ldots ,c_i,\ldots }\left( \ldots
,p_i,\ldots \right) \mid _{_{\left\{ p_i\right\} =\left\{ p_i^{\prime
}\right\} }}
\end{equation}

Finally, in order to take into account the gauge invariance with respect to
the second gluon in the process one should replace $l_2$ by $l_2^{\prime }$
where $l_2^{\prime }p_2^{\prime }=0$ and a difference between $l_2$ and $
l_2^{\prime }$ is order of $O\left( k\right) $. Then $l_2^{\prime }$ is
reduced to $l_2^{\bot ^{\prime }}$ where $l_2^{\prime }=B\left( l_2^{\prime
}\right) p_2^{\prime }+$ $l_2^{\bot ^{\prime }}$ and $B\left( p_2^{\prime
}\right) =1,\;l_2^{\bot ^{\prime }}p_2^{\prime }=0$. The notation $\bot
^{\prime }$ emphasizes the transversality with respect to $p_2^{\prime }$.
In this way we obtain the expression for the total inelastic amplitude of
the gluon production with small momentum $k$ in terms of the elastic
amplitude.

\begin{equation}
\label{11}
\begin{array}{c}
A_g^{tot\ a,\ a_1,a_2,a_3,a_4}\left( l,k;p_1;l_2,p_2;p_3;p_4\right) = \\
\\
=+ig\sum_i\sum_{c_i}f_{c_i}^{a\;a_i}
\frac{lp_i}{kp_i}\left( l_{2\ \nu }^{\;\bot ^{\prime }}A^{\nu ,\ \ldots
,c_{i,\ldots }}\left( \ldots ,p_i^{\prime }{}_{,\ldots }\right) +\right.
\qquad \qquad \qquad \quad \\  \\
\left. +\sum_j\xi _j\frac \partial {\partial p_j}A^{\nu ,\ \ldots
,c_i,\ldots }\left( \ldots ,p_i{}_{,\ldots }\right) \mid _{\left\{
p_i\right\} =\left\{ p_i^{\prime }\right\} }\right) + \\
\\
+ig\sum_i\sum_{c_i}f_{c_i}^{a\;a_i}\left( l_{2\ \nu }^{\;\bot ^{\prime
}}D_iA^{\nu ,\ \ldots ,c_{i,\ldots }}\left( \ldots ,p_i,\ldots \right)
\right) \mid _{_{\left\{ p_i\right\} =\left\{ p_i^{\prime }\right\} }}+\quad
\\
\\
+ig\sum_cf_c^{a\;a_2}\frac{\left( \left( l_2^{\;\bot ^{\prime }}l\right)
k_{\ \nu }-l_{\ \nu }\left( l_2^{\;\,\bot ^{\prime }}k\right) \right) }{kp_2}
A^{\nu ,\ a_1,c,a_3,a_4}\left( \ldots ,p_2^{\prime },\ldots \right)
\end{array}
\end{equation}

where $D_i\left( k\right) =\frac{p_i}{kp_i}\left( k\frac \partial {\partial
p_i}\right) -\frac \partial {\partial p_i}$. Then the property $k\cdot
D_i\left( k\right) =0$ and charge conservation expressed as $
\sum_i\sum_{c_i}f_{c_i}^{a\;a_i}A^{\nu ,\ \ldots ,c_{i,\ldots }}\left(
\ldots ,p_i^{\prime }{}_{,\ldots }\right) =0$ imply the required
gauge-invariance properties $A_g^{tot\ a,\ a_1,a_2,a_3,a_4}\left(
l=k,\;k;\ldots ;\ldots \right) =0$. The dependence on $\xi _j$ is decorative
and has to fall out in the final result.

The first term in the expression $\left( \ref{11}\right) $ involves terms of
order of $O\left( \omega ^{-1}\right) $  while other terms
are the corrections of order of  $O\left( \omega ^0\right) $ where $\omega$
is the boson frequency in the c.m. system. The last term in (\ref{11}) can
be treated as the contribution of a color anomalous magnetic moment.

The region of applicability of the Low formula can be extended \cite{3} to
the high-energy region where the momentum of the radiated gauge particle is
not small. Namely, it is sufficient to require that the emission of the
gauge boson does not change the kinematics of the basic elastic process. Let
us consider the basic elastic process of scattering in the Regge kinematics
(see Fig.5) where $2p_1\cdot p_2=s\ \gg \ m_{char}^2$ and $\left(
p_1+p_4\right) ^2=t\sim m_{char}^2$ , so that $\frac
ts\rightarrow 0.$ Going to the inelastic amplitude (see Fig.6),
we introduce a Sudakov decomposition of the momentum of the
radiated gluon

\begin{equation}
\label{12}k=\alpha p_2+\beta
p_1+k^{\bot }
\end{equation}

where $k^{\bot }p_1=k^{\bot }p_2=0,\;s\alpha \beta =k^{\bot \;2}$. Then the
requirement to keep the kinematics of the basic process can be written as
the following condition on the parameters $\alpha ,\;\beta $

\begin{equation}
\label{13}\alpha \;\ll \;1\quad ,\quad \beta \;\ll \;1\quad ,\quad \left|
k^{\bot }\right| \;\ll \;m_{ch}
\end{equation}

that is the fraction of energy which the gluon takes from the incoming
particles is small and influence of the bremsstrahlung radiation onto the
momentum transfer in the basic process is negligible.
\begin{equation}
\label{14}
\begin{array}{c}
s^{\prime }-s=\left( p_1+p_3\right) ^2-\left( p_1+p_2\right) ^2\approx
-\left( \alpha +\beta \right) s\ll s \\
\\
t_2-t_1=2qk^{\bot }\ll t
\end{array}
\end{equation}

We will look more carefully at this region below. The starting point of the
further consideration is the elastic amplitude for the collision of three
scalar and one vector particles (see Fig.5) in the Regge kinematics. In this
case the polarization vector of the gluon can be decomposed into its
longitudinal and transverse components with respect to momenta $p_1$ and $
p_2 $

\begin{equation}
\label{15}l_2=2\frac{\left( l_2p_1\right) }sp_2+l_2^{\bot }
\end{equation}

where $l_2^{\bot }p_1=l_2^{\bot }p_2=0$. The elastic amplitude (in Fig.5) is
factorized in its spin and unitary isospin indices according to a definite
t-channel group representation

\begin{equation}
\label{16}A_{2\rightarrow 1+g\;\;\;a_1\;a_2}^{\quad \quad \quad
a_4\;a_3}\left( s,t\right) =\left( l_2^{\bot }q\right) A_{2\rightarrow
2\;\;\;a_1\;a_2}^{\quad \quad a_4\;a_3}\left( s,t\right)
\end{equation}

\begin{equation}
\label{17}
\begin{array}{c}
A_{2\rightarrow 2\;\;\;a_1\;a_2}^{\quad \quad a_4\;a_3}\left( s,t\right)
=\sum_c\gamma _{a_1}^{a_4}\left( c\right) \gamma _{a_2}^{a_3}\left( c\right)
A\left( s,t\right) = \\
\\
\sum_c\gamma _{a_1}^{a_4}\left( c\right) \gamma _{a_2}^{a_3}\left( c\right)
\beta \left( t\right) \left( \left( -\frac S{m_{ch}}\right) ^{\alpha \left(
t\right) }+\left( \frac S{m_{ch}}\right) ^{\alpha \left( t\right) }\right)
\end{array}
\end{equation}

where $q=p_1+p_4=-p_2-p_3$ and $a_i$ are the color indices of initial and
final particles, $c$ is an isotopic index of a reggeon, $\gamma
_{a_1}^{a_4}\left( c\right) $ are Clebsch-Gordon coefficients. $A
_{2\rightarrow 2\;\;\;a_1\;a_2}^{\quad \quad a_4\;a_3}\left( s,t\right) $ is
the elastic amplitude for the collision of the scalar particles in the Regge
kinematics. Below we use relations
$$
\sum_{c,d,e}\gamma _{a_2}^{a_3}\left( c\right) \left(
f_d^{a\;a_4}+f_{a_1}^{a\;e}\right) \gamma _e^d\left( c\right)
=\sum_{c,d}\gamma _{a_1}^{a_4}\left( c\right) T_{cd}^a\gamma
_{a_2}^{a_3}\left( d\right)
$$
$$
\sum_{c,d,e}\gamma _{a_2}^{a_3}\left( c\right) \left(
f_d^{a\;a_3}+f_{a_2}^{a\;e}\right) \gamma _e^d\left( c\right)
=-\sum_{c,d}\gamma _{a_1}^{a_4}\left( c\right) T_{cd}^a\gamma
_{a_2}^{a_3}\left( d\right)
$$

or

\begin{equation}
\label{18}\left( T_1+T_4\right) \gamma \left( c\right) =T\gamma \left(
c\right) \quad ;\quad \gamma \left( c\right) \left( T_2+T_3\right) =-\gamma
\left( c\right) T
\end{equation}

where the generators $T_i$ in Eq. $\left( \ref{18}\right) $ act on the
unitary spin indices of the i-th particle $\left( T_i\right)
_c^a=f_{a_i\;c}^{\;a}$ and due to the gauge invariance
\begin{equation}
\label{19}\left( T_1+T_2+T_3+T_4\right) A_{2\rightarrow 2}\left( s,t\right)
=0,
\end{equation}
while the generator $T$ acts on the unitary indices of the reggeon. For
present purposes a representation of the $\xi _i\left( k\right) $ is defined
by conservation of momenta and gauge invariance. We can obtain the $\xi
_i\left( k\right) $ in the form $\xi _{1,2}=\frac{s_{1,2}+t_{1,2}}sp_{1,2}+
\frac{k^{\bot }}2$ , $\xi _3=\xi _4=0$ or, in other words, the physically
realizable set of on-mass-shell momenta \{$p_i^{\prime }$\} is given by

\begin{equation}
\label{20}
\begin{array}{c}
p_1^{\prime }=p_1\left( 1-
\frac{s_1+t_1}s\right) +\frac{k^{\bot }}2 \\ p_2^{\prime }=p_2\left( 1-
\frac{s_2+t_2}s\right) +\frac{k^{\bot }}2 \\ p_3^{\prime }=p_3\quad ,\quad
p_4^{\prime }=p_4 \\
q=p_1^{\prime }+p_4^{\prime }=-p_2^{\prime }-p_3^{\prime }
\end{array}
\end{equation}

The contributions of the $\xi $ -terms are of order of $O\left( k\right) $
in the bremsstrahlung amplitude and falls out. In this way for the kinematics

\begin{equation}
\label{21}
\begin{array}{c}
s_2\sim s_{2^{\prime }}\ll t_1\sim t_2\sim m_{ch}^2 \\
\\
s_1\sim s\gg m_{ch}^2
\end{array}
\end{equation}

we take into account the requirement (\ref{13}) and continue the Low
expression of the inelastic amplitude to this region,

$$
\begin{array}{c}
A_g^{tot\ a,\ a_1,a_2,a_3,a_4}\left( l;k;p_1;l_2,p_2;p_3;p_4\right) =
\end{array}
$$
$$
\begin{array}{c}
I \\
ig\left\{ -2\frac{lp_1}{s_1}\sum_{c,d}\gamma _{a_1}^{a_4}\left( c\right)
T_{dc}^a\gamma _{a_1}^{a_3}\left( d\right) \left( l_2^{\;\bot ^{\prime
}}q\right) A\left( s,t\right) \right.
\end{array}
$$
$$
\begin{array}{c}
II \\
-2\frac{lp_2}{s_2}\sum_{c,d}\gamma _{a_1}^{a_5}\left( c\right) \left(
T_2\right) ^{a\;d}\gamma _d^{a_3}\left( c\right) \left( l_2^{\;\bot ^{\prime
}}q\right) A\left( s,t\right)
\end{array}
$$
$$
\begin{array}{c}
III \\
+2\frac{lp_3}{s_{2^{\prime }}}\sum_{c,d}\gamma _{a_1}^{a_5}\left( c\right)
\left( T_3\right) _d^{a_4}\gamma _{a_2}^d\left( c\right) \left( l_2^{\;\bot
^{\prime }}q\right) A\left( s,t\right)
\end{array}
$$
\begin{equation}
\label{22}
\begin{array}{c}
IV \\
+\left[ -\left(
\frac{lp_1}{s_1}\left( l_2^{\;\bot ^{\prime }}k\right) +\frac 12\left(
l_2^{\;\bot ^{\prime }}l\right) \right) +\left( l_2^{\;\bot ^{\prime
}}q\right) \left( lB_1\right) \frac \partial {\partial t}\right] \times \\
\\
\times \sum_{c,d}\gamma _{a_1}^{a_4}\left( c\right) T_{dc}^a\gamma
_{a_1}^{a_3}\left( d\right) \left( l_2^{\;\bot ^{\prime }}q\right) A\left(
s,t\right)
\end{array}
\end{equation}
$$
\begin{array}{c}
V \\
+\left[ \left(
\frac{lp_2}{s_2}\left( l_2^{\;\bot ^{\prime }}k\right) +\frac 12\left(
l_2^{\;\bot ^{\prime }}l\right) \right) +2\frac 1{s_2}\left( \left( l\left(
l_2^{\;\bot ^{\prime }}k\right) -\left( l_2^{\;\bot ^{\prime }}l\right)
k\right) _\nu q^\nu \right) \right. \\  \\
\left. -\left( l_2^{\;\bot ^{\prime }}q\right) \left( lB_2\right) \frac
\partial {\partial t}\right] \times \\
\\
\times \sum_{c,d}\gamma _{a_1}^{a_4}\left( c\right) \left( T_2\right)
^{a\;d}\gamma _d^{a_3}\left( c\right) \left( l_2^{\;\bot ^{\prime }}q\right)
A\left( s,t\right)
\end{array}
$$
$$
\begin{array}{c}
VI \\
+\left[ \left( -
\frac{lp_3}{s_{2^{\prime }}}\left( l_2^{\;\bot ^{\prime }}k\right) +\frac
12\left( l_2^{\;\bot ^{\prime }}l_4\right) \right) -\left( l_2^{\;\bot
^{\prime }}q\right) \left( l_4B_{2^{\prime }}\right) \frac \partial
{\partial t}\right] \times \\  \\
\times \sum_{c,d}\gamma _{a_1}^{a_4}\left( c\right) \left( T_3\right)
_d^a\gamma _{a_2}^d\left( c\right) \left( l_2^{\;\bot ^{\prime }}q\right)
A\left( s,t\right)
\end{array}
$$

where

\begin{equation}
\label{23}
\begin{array}{c}
B_1=\left( t_1-t_2\right)
\frac{p_1}{s_2}-\frac{q_1+q_2}2\qquad ,\qquad kB_1=0 \\ B_2=\left(
t_1-t_2\right)
\frac{p_2}{s_2}-\frac{q_1+q_2}2\qquad ,\qquad kB_2=0 \\ B_{2^{\prime
}}=-\left( t_1-t_2\right)
\frac{p_3}{s_{2^{\prime }}}-\frac{q_1+q_2}2\qquad ,\qquad kB_{2^{\prime }}=0
\\ t_i=q_i^2\ ,\ q_1=p_1+p_4\ ,\ q_2=p_2+p_3
\end{array}
\end{equation}

In order to derive $\left( \ref{22}\right) $ we have used that in the
kinematics in $\left( \ref{21}\right) $ one has
\begin{equation}
\label{23a}
\begin{array}{c}
D_1\left( k\right) A\left( s,t\right) =B_1A\left( s,t\right) \approx
D_4\left( k\right) A\left( s,t\right) \\
\\
D_2\left( k\right) A\left( s,t\right) =B_2A\left( s,t\right) \quad
D_{2^{\prime }}\left( k\right) A\left( s,t\right) =B_{2^{\prime }}A\left(
s,t\right) \\
\\
B_1\approx B_{1^{\prime }}=B_5\qquad s_2B_2=s_{2^{\prime }}B_{2^{\prime }}
\end{array}
\end{equation}

Last three terms $IV,V$ and $VI$ in Eq.$\left( \ref{22}\right) $ are
corrections of order of $\left( k^{\bot }\right) ^0$ to the first three
terms which are of the order $\left( k^{\bot }\right) ^{-1}$. Unlike the
scalar particle collisions, Eq.$\left( \ref{22}\right) $ contains the
additional terms $\sim \left( k^{\bot }\right) ^0$ related to the non-zero
spin particles.  In the $s_2$ -channel the corrections $\sim \left(
\frac{lp_2}{s_2}\left( l_2^{\;\bot ^{\prime }}k\right) +\frac 12\left(
l_2^{\;\bot ^{\prime }}l\right) \right) +2\frac 1{s_2}\left( \left( l\left(
l_2^{\;\bot ^{\prime }}k\right) -\left( l_2^{\;\bot ^{\prime }}l\right)
k\right) _\nu q^\nu \right) $ related to the  anomalous color quadrupole
electric and magnetic dipole moments of the gluon. In the $s_1$ and $
s_{2^{\prime }}$ -channels the corrections of order of $\left( k^{\bot
}\right) ^0$ like terms $\sim \left( -\frac{lp_i}{s_i}\left( l_2^{\;\bot
^{\prime }}k\right) +\frac 12l_2^{\;\bot ^{\prime }}l\right) $ can be
treated as induced vertices for two color scalar particles with two
vectors.

Let us consider the expression in $\left( \ref{22}\right) $ in the limit $
s_2\rightarrow \infty $ that is in the multi-Regge kinematics

\begin{equation}
\label{24}
\begin{array}{c}
s_1\gg m_{ch}^2\quad ,\quad s_2\gg m_{ch}^2\quad ,\quad
\frac{s_1s_2}s=\overrightarrow{k}_{\bot }^2\ll m_{ch}^2 \\  \\
t_1,t_2\sim m_{ch}^2
\end{array}
\end{equation}

In this case for $k_{\bot }\ll m_{ch}$ the amplitude in $\left( \ref{22}
\right) $ takes the form
$$
\left. A_{2\rightarrow 2+g}^{\ a,\
a_1,a_2,a_3,a_4}(s,s_1,s_2,t_1,t_2)\right| _{k_{\bot }\rightarrow
0}=\;g^3\;\;\times
$$
\begin{equation}
\label{24a}l_4^\mu \left[ 2\left( \frac{p_2}{s_2}-\frac{p_1}{s_1}\right)
\left( l_2^{\,\bot ^{\prime }}q\right) -l_2^{\,\bot ^{\prime }}-\left( \frac{
p_2}{s_2}+\frac{p_1}{s_1}\right) \left( l_2^{\,\bot ^{\prime }}k\right)
+(B_1+B_2)\frac \partial {\partial t}\right] _\mu \times
\end{equation}
$$
\times\sum_{i,j}\gamma _{a_1}^{a_4}\left( i\right) T_{ji}^a\gamma
_{a_1}^{a_3}\left( j\right) \left( l_2^{\;\bot ^{\prime }}q\right) A\left(
s,t\right) \quad .
$$

The first term in (\ref{24a}) is the leading pole Gribov term of order $
\frac 1{\left| k^{\bot }\right| }$ and other are the corrections of order $
\left| k^{\bot }\right| ^0$. In this formula contribution of the magnetic
moment is suppressed as $O\left( \frac{\overrightarrow{k^{\bot }}^2}{s_2}
\right) $. But there are terms $\sim -l_2^{\,\bot ^{\prime }}-\left( \frac{
p_2}{s_2}+\frac{p_1}{s_1}\right) \left( l_2^{\,\bot ^{\prime }}k\right) $
and this expression does not agree with the amplitude factorization which
must be in the multi-Regge kinematics. In order to correct the situation it
is necessary to take into account some additional contributions. Below we
represent the correct amplitude and explain the mechanism of appearance of
this contributions using the bosonic string as guide.

\section{Contribution of intermediate excited states in the multi-Regge
kinematics.}

In this section we will try to take into account the contribution of
intermediate excited states in the $s_1,s_{1^{\prime }},s_2$ and $
s_{2^{\prime }}$ channels in addition to the Low formula above, which
describes the contribution of the ground state in these channels. For the
present case the ground states are considered to be scalar in the $
s_1,s_{1^{\prime }},s_{2^{\prime }}$ channels and vector in the $s_2$
channel. It is important to emphasize that the contribution from excited
intermediate states which is small $\left( \sim \omega \right) $ due to
gauge invariance \cite{3}. However, in the regions of the large $s_i$ it
turns out is of order $\left| k^{\bot }\right| ^0$, that is of the same
order as of the ordinary correction of the ground states to the Low formula.
The necessity of the inelastic contributions follows from the correct
analytic behavior for the scattering amplitude \cite{5}. Taking into account
the excited states we have to relax the condition $s_i\sim 0$. The simplest
way is to consider the production of massless particles with small
transverse momenta in the multi-Regge kinematics in such a model which would
allow us to take into account excited states.

As discussed earlier, the leading terms $\sim O\left( \frac 1{\left| k^{\bot
}\right| }\right) $ in the Low formula describe correctly the gluon
bremsstrahlung amplitude in regions $\left( \ref{13}\right) $ and $\left(
\ref{24}\right) $. As shown in Ref.\cite{5} the correction related to
excited states is fixed by general analyticity in the channels $s_1,$ and $
s_2$ up to a constant which cannot be obtained from general consideration.
Calculation of this constant was illustrated by an example of string bosonic
model, when the gauge particle was produced in scattering of four scalar
particles \cite{4} and also of four massless vector particles \cite{6}.
Since we would like to establish the dependence of the gluon bremsstrahlung
amplitude on the nature of the external particles then it is interesting to
consider the mixed case of particles with different spin. In the multi-Regge
kinematics the calculation is rather simple, furthermore, it also allows us
to control further calculations in other region.

In what follows we conform to the bosonic string model to illustrate the
further consideration of the problem. The first step is to consider the
elastic string amplitude of the scattering of the scalar particles and a
vector particle in the Regge kinematics. In this process tachyons which are
ground states in the string play a role of the scalar particles while the
vector particle is an excited state of the open string. In the Born
approximation we have (see Fig.5)

\begin{equation}
\label{25}A_{2\rightarrow 1+g}(s,t,u)\;\sim \;\int
\prod_{i=1}^4dx_i\prod_{i<j}|x_i-x_j|^{-2\alpha ^{\prime }p_ip_j}V^1(x),
\end{equation}
where $p_1,p_2,p_3$ and $p_4$ are the momenta of external on-shell particles
$p_i^2=p_{i_0}^2-\vec p_i^2=-\frac 1{\alpha ^{\prime }}$ $\left( i\neq
2\right) $ and $p_2^2=0$. We choose independent scalar invariants $t=\left(
p_1+p_4\right) ^2$ and $s=2p_1p_2$. Other scalar products are linear
combinations of the $s$ and $t$ ; $\alpha ^{\prime }$ is a slope of the
Regge trajectory $\alpha (t)=1+\alpha ^{\prime }t$. The vertex $V^1(x)$
depends on polarization vector $l_2^\mu $, momenta $p_i^\mu $ $\left( i\neq
2\right) $ and Koba-Nielsen variables $x_i$.

\begin{equation}
\label{26}V^1(x)=\sum_{m\neq 2}\frac{l_2p_m}{x_m-x_2}
\end{equation}

It is not difficult to verify invariance of Eq.$\left( \ref{26}\right) $
with respect to gauge transformation $l_i(p_i)\to l_i(p_i)+cp_i$, taking
into account the identity

\begin{equation}
\label{27}p_2\,\sum_{i\neq 2}\,\frac{(p_il_2)}{x_i-x_2}\;=\;\frac \partial
{\partial x_2}\;\sum_{i<j}2p_ip_j\ln |x_i-x_j|.
\end{equation}

The vertex $V^1(x)$ and expression $\left( \ref{25}\right) $ are invariant
with respect to the M\"obius group
\begin{equation}
\label{28}x_i^{\prime }\;\;=\;\frac{(ax_i+b)}{(cx_i+d)},\qquad ad-bc\;=\;1
\end{equation}
and we have to take into account the factor $|x_{13}x_{34}x_{41}|$ which
gives a volume of this group to obtain the correct expression of the
amplitude
\begin{equation}
\label{29}A^4(s,t)\;\sim \int \prod_{i=1}^4dx_i\left| \frac{x_{21}x_{34}}{
x_{31}x_{42}}\right| ^{-\alpha ^{\prime }s}\left| \frac{x_{41}x_{32}}{
x_{31}x_{42}}\right| ^{-\alpha ^{\prime }t}V^1(x)|x_{13}x_{34}x_{41}|,
\end{equation}
where $x_{ij}=x_i-x_j$. Thanks to $\left( \ref{28}\right),$ we can fix three
variables of integration to be
\begin{equation}
\label{30}x_1=\;0,\qquad x_2=\;x,\qquad x_3=\;1,\qquad x_4=\;\infty .
\end{equation}

The Yang-Mills color group is introduced simply by multiplying every string
diagrams by the corresponding Chan-Paton factor \cite{8} according to the
order of external particles. In such a case, different regions of
integration are combined in the amplitude, so that every one enters with its
corresponding weight
\begin{equation}
\label{31}
\begin{array}{ccc}
C_{a_1a_2a_3a_4} & = & \frac 12\,tr\left( \gamma _{a_1}\gamma _{a_2}\gamma
_{a_3}\gamma _{a_4}\right) ,\;\;tr\,\gamma _i\gamma _j=\,\delta _{ij} \\
&  &  \\
C_{a_1a_2\ldots a_n} & = & (-1)^nC_{a_na_{n-1}\ldots a_1},
\end{array}
\end{equation}
where $\gamma _i$ are matrices of the fundamental representation of color
$SU(N)$ group.

Thus, after the normalization of the amplitude we get
\begin{equation}
\label{32}
\begin{array}{ccc}
A_{2\rightarrow 1+g}(s,t) & = & g^2\left( C_{a_2a_1a_3a_4}\int_{-\infty
}^0+\;C_{a_1a_2a_3a_4}\int_0^1+\;C_{a_1a_3a_2a_4}\int_1^\infty \right)
|x|^{-\alpha (s)-1} \\
&  &  \\
& \times & |1-x|^{-\alpha (t)-1}\;V^1(x)\,dx.
\end{array}
\end{equation}
In the case of the string model we have $m_{ch}\sim \sqrt{\frac 1{\alpha
^{\prime }}}$ and in the Regge asymptotics
\begin{equation}
\label{33}s\;\gg \;\frac 1{\alpha ^{\prime }},\qquad t\;\sim \;\frac
1{\alpha ^{\prime }},
\end{equation}
the essential region of integration is $|1-x|=\epsilon \sim 1/\alpha
^{\prime }s$ and the first term in Eq.(\ref{32}) turns to be asymptotically
small.

In the multi-Regge kinematics it is convenient to expand the polarization
vector into longitudinal and transverse parts with respect to the vectors $
p_1$ and $p_2$ (see$\left( \ref{15}\right) $). The main asymptotic
contribution in the amplitude takes the form
\begin{equation}
\label{34}A_{2\rightarrow 1+g}^{a_1a_2a_3a_4}(s,t)\;=\;g^2\left( l_2^{\bot
}q\right) \left( C_{a_1a_2a_3a_4}\int_0^\infty
+\;C_{a_1a_3a_2a_4}\int_{-\infty }^0\right) \,e^{-\alpha ^{\prime }s\epsilon
}\epsilon ^{-\alpha ^{\prime }t}\frac{d\epsilon }{\epsilon ^2},
\end{equation}

where $q=p_1+p_4=-p_2-p_3$ (see Fig.5).

After the integration and the analytical continuation of this expression we
obtain the elastic amplitude which has the factorized form
\begin{equation}
\label{35}A_{2\rightarrow 1+g}^{_{a_1a_2a_3a_4}}(s,t)\;=\left( l_2^{\bot
}q\right) A_{2\rightarrow 2}^{_{a_1a_2a_3a_4}}(s,t),
\end{equation}

where
\begin{equation}
\label{36}A_{2\rightarrow 2}^{_{a_1a_2a_3a_4}}(s,t)\;=\;g^2\left( (-\alpha
^{\prime }s)^{\alpha (t)}C_{a_1a_2a_3a_4}+(\alpha ^{\prime }s)^{\alpha
(t)}C_{a_1a_3a_2a_4}\right) \,\Gamma (-\alpha (t)).
\end{equation}
$A_{2\rightarrow 2}(s,t)$ is the elastic scattering tachyon amplitude
\cite{4}, $\Gamma (x)$ is the Gamma-function.

Now we consider the gluon production amplitude in the Koba-Nielsen
representation (see Fig.7)

\begin{equation}
\label{37}A_{2\rightarrow 1+g+g}(s,s_1,s_2,t_1,t_2)\,\sim \int
\prod_{i=1}^5dx_i\prod_{i<j}|x_i-x_j|^{-2\alpha ^{\prime }p_ip_j}V^2(x_i).
\end{equation}

Here momenta $p_i$ of external particles are on mass shell $
p_2^2=p_4^2=0,p_1^2=p_3^2=p_5^2=-\frac 1{\alpha ^{\prime }}$. We choose as
independent the following scalar invariants
\begin{equation}
\label{38}
\begin{array}{c}
s=2p_1p_2,\quad s_1=-2p_1p_4,\quad s_2=-2p_2p_4, \\
\\
t_1=2p_1p_5-\frac 2{\alpha ^{\prime }},\quad t_2=2p_2p_3-\frac 1{\alpha
^{\prime }}.
\end{array}
\end{equation}
All other invariants can be expressed in terms of the above ones as
$$
2p_1p_3\;=\;-s+s_1-t_1,\qquad 2p_2p_5\;=\;-s+s_2-t_2-\frac 1{\alpha ^{\prime
}}
$$
\begin{equation}
\label{39}2p_3p_4\equiv s_{2^{\prime }}=s_2+t_1-t_2,\quad
2p_3p_5=s-s_1-s_2-\frac 1{\alpha ^{\prime }},
\end{equation}
$$
2p_4p_5\equiv s_{1^{\prime }}=s_1-t_1+t_2.
$$
The vertex $V^2(x_i)$ of the emission of the two vector particles is given
by \cite{6,9}.

\begin{equation}
\label{40}V^2\left( x_i\right) =-\frac{l_2l_4}{\left( x_2-x_4\right) ^2}
+2\alpha ^{\prime }\sum_2^5\sum_4^5
\end{equation}

where
$$
\sum_m^5=\sum_{l\neq m}^5\frac{l_mp_l}{x_l-x_m}\ .
$$

We also add the factor $\left| x_{15}x_{12}x_{52}\right| $ to cancel the
infinite volume of the M\"obius group and fix three Koba-Nielsen variables
$$
x_1=0,\;x_2=1,\;x_3=x,\;x_4=y,\;x_5=\infty \;
$$
and then rewrite the expression of the amplitude as follows

\begin{equation}
\label{42}
\begin{array}{r}
A_{2\rightarrow 1+g+g}(s,s_1,s_2,t_1,t_2)\,\sim \int dxdy\left| \frac
1x\right| ^{-\alpha ^{\prime }s}\left| \frac xy\right| ^{-\alpha ^{\prime
}s_1}\left|
\frac{x-y}{1-y}\right| ^{-\alpha ^{\prime }s_2}\times \\  \\
\times \left|
\frac{x-y}x\right| ^{-\alpha ^{\prime }t_1}\left| \frac{1-x}{x-y}\right|
^{-\alpha ^{\prime }t_2}\left| \frac x{\left( 1-x\right) ^2}\right| \times
\\ \\
\times \left( -\frac{l_2l_4}{y^2}+2\alpha ^{\prime }l_2^\mu \left( p_2+\frac{
p_3}x+\frac{p_4}y\right) _\mu \;l_4^\nu \left( -\frac{p_1}y+\frac{p_2}{1-y}+
\frac{p_3}{x-y}\right) _\nu \right)
\end{array}
\end{equation}

The fourth particle are considered as the radiated one so we denote $k=p_4$
and $l=l_4$. Similarly to the cases of scattering of tachyons \cite{4} or
gluons \cite{6} we first estimate the expression in $\left( \ref{42}\right) $
in the limit $\overrightarrow{k_{\bot }}^2\rightarrow 0$ and then take the
limit $\overrightarrow{k_{\bot }}\rightarrow 0$.

In the multi-Regge kinematics in $\left( \ref{24}\right) $ the main
contribution gives the region
\begin{equation}
\label{43}
\begin{array}{c}
\left.
\begin{array}{c}
\frac xy-1\equiv \epsilon _1\sim \frac 1{\alpha ^{\prime }s_1}\rightarrow 0
\\
\\
\frac{x-y}{1-y}\equiv \epsilon _2\sim \frac 1{\alpha ^{\prime
}s_2}\rightarrow 0
\end{array}
\right\} \qquad \Rightarrow \qquad
\begin{array}{c}
x\approx 1+\epsilon _1\epsilon _2 \\
\\
y\approx 1-\epsilon _1
\end{array}
\qquad .
\end{array}
\end{equation}

Thus, for the new variables the representation of the amplitude in $\left(
\ref{42}\right) $ is to be divided into four pieces which correspond to
positive or negative $\epsilon _1,\epsilon _2$ . In each piece the values of
$s,s_1$ and $s_2$ have to be chosen positive or negative in such a way, that
integrals converge. After integration one has to continue them to the
physical region of the s-channel (or another channel) where $s,s_1$ and $s_2$
take positive values. In addition we take into account their color
Chan-Paton factors $\left( \ref{31}\right) $ (see Fig.8) and introduce new
variables $x_1,\;x_2$
$$
x_1=\alpha ^{\prime }s_1\epsilon _1\quad ,\quad x_2=\alpha ^{\prime
}s_2\epsilon _2\ .
$$
After extracting the factors $\left( \alpha ^{\prime }s_1\right) ^{\alpha
\left( t_1\right) }$ and $\left( \alpha ^{\prime }s_2\right) ^{\alpha \left(
t_2\right) }$, the remaining integral depends on the invariants $s,s_1$ and $
s_2$ only via the combination $\Lambda =\frac s{\alpha ^{\prime
}s_1s_2}=-\frac 1{\alpha ^{\prime }\overrightarrow{k_{\bot }}^2}$. For small
$\overrightarrow{k_{\bot }}^2\rightarrow 0$ and therefore for large $\Lambda
\rightarrow \infty $ the main contribution comes from the regions $x_1\sim
\frac 1\Lambda \;,\;x_2\sim 1$ and $x_2\sim \frac 1\Lambda \;,\;x_1\sim 1$.
In doing so, we obtain the inelastic amplitude (see Fig.7) in the
multi-Regge kinematics, which is factorized in its spin indices
\begin{equation}
\label{44}\left. A_{2\rightarrow
1+g+g}^{a_1a_2a_3a_4a_5}(s,s_1,s_2,t_1,t_2)\right| _{\overrightarrow{k_{\bot
}}^2\rightarrow 0}=\left( l_2^{\bot }q_2\right) \left. A_{2\rightarrow
2+g}^{a_1a_2a_3a_4a_5}(s,s_1,s_2,t_1,t_2)\right| _{\overrightarrow{k_{\bot }}
^2\rightarrow 0}
\end{equation}

where $\left. A_{2\rightarrow 2+g}(s,s_1,s_2,t_1,t_2)\right| _{
\overrightarrow{k_{\bot }}^2\rightarrow 0}$ is the gluon production
amplitude in the case of the collision involving only tachyons \cite{4}
$$
\left. A_{2\rightarrow 2+g}^{a_1a_2a_3a_4a_5}(s,s_1,s_2,t_1,t_2)\right| _{
\overrightarrow{k_{\bot }}^2\rightarrow 0}=\;\alpha ^{\prime }g^3\;\;\times
$$
$$
\;\left\{ C_{a_1a_2a_3a_4a_5}\left( (-\alpha ^{\prime }s)^{\alpha
(t_1)}(-\alpha ^{\prime }s_2)^{\alpha (t_2)-\alpha (t_1)}J_2+(-\alpha
^{\prime }s)^{\alpha (t_2)}(-\alpha ^{\prime }s_1)^{\alpha (t_1)-\alpha
(t_2)}J_1\right) \right.
$$
\begin{equation}
\label{45}+\;C_{a_1a_4a_2a_3a_5}\left( (-\alpha ^{\prime }s)^{\alpha
(t_1)}(\alpha ^{\prime }s_2)^{\alpha (t_2)-\alpha (t_1)}J_2+(-\alpha
^{\prime }s)^{\alpha (t_2)}(\alpha ^{\prime }s_2)^{\alpha (t_1)-\alpha
(t_2)}J_1\right)
\end{equation}
$$
+\;\;C_{a_1a_3a_2a_4a_5}\left( (\alpha ^{\prime }s)^{\alpha (t_1)}(\alpha
^{\prime }s_2)^{\alpha (t_2)-\alpha (t_1)}J_2+(\alpha ^{\prime }s)^{\alpha
(t_2)}(-\alpha ^{\prime }s_1)^{\alpha (t_1)-\alpha (t_2)}J_1\right)
$$
$$
\left. +\;C_{a_1a_4a_3a_2a_5}\left( (\alpha ^{\prime }s)^{\alpha
(t_1)}(\alpha ^{\prime }s_2)^{\alpha (t_2)-\alpha (t_1)}J_2+(\alpha ^{\prime
}s)^{\alpha (t_2)}(\alpha ^{\prime }s_1)^{\alpha (t_1)-\alpha
(t_2)}J_1\right) \right\}
$$

where

\begin{equation}
\label{46}
\begin{array}{lll}
J_1 & = & \Gamma (-\alpha (t_2))\Gamma (\alpha (t_2)-\alpha (t_1))B_\mu
^1l^\mu \ . \\
&  &  \\
J_2 & = & \Gamma (-\alpha (t_1))\Gamma (\alpha (t_1)-\alpha (t_2))B_\mu
^2l^\mu .
\end{array}
\end{equation}

The amplitude in (\ref{44}) has the correct analytical structure and the
simultaneous singularities in the $s_1$ and $s_2$ channels are absent.
Consider now the result in $\left( \ref{46}\right) $ in the region
\begin{equation}
\label{47}\left| \overrightarrow{k_{\bot }}\right| \ll m_{ch}\quad .
\end{equation}
In this case $m_{ch}\sim q_i\sim \frac 1{\sqrt{\alpha ^{\prime }}}$ and in
the physical region we have simultaneously $\frac 1\Lambda \rightarrow 0$
and $\alpha ^{\prime }\left( t_1-t_2\right) =\alpha ^{\prime }\left(
q_1+q_2\right) \left( q_1-q_2\right) \sim \alpha ^{\prime }\left( qk_{\bot
}\right) \rightarrow 0$. We want to obtain an expression for the total
inelastic amplitude of the gluon production with small momentum $k_{\bot }$
in terms of elastic amplitude $\left( \ref{35}\right) $. For this purpose,
as discussed earlier, we have to express the momenta and the polarization
vector of the inelastic amplitude in terms of the corresponding values in
the elastic amplitude. But we have to be careful to keep the gauge
invariance and the conservation of momenta. In other words we take the
physically realizable set of on-mass-shell momenta \{$p_i^{\prime }$\} (see
Eq.$\left( \ref{20}\right) $) and the polarization vector $l_2^{\prime }$
for the elastic amplitude
$$
\begin{array}{lll}
l_2^{\prime } & = & l_2-\frac{l_2^{\bot }k}s\,p_1+\frac{l_2p_1}s\,k^{\bot }
\end{array}
$$
or, in terms of the transverse component
$$
\begin{array}{lll}
l_2^{\perp } & = & l{}_2^{\bot ^{\prime }}-\frac{(l_2{}^{\perp ^{\prime }}k)}
s\,p_1^{\prime }-\frac{(l_2^{\perp ^{\prime }}k)}s\,p_2^{\prime }
\end{array}
$$

Expanding Eq.(\ref{45}) in $k^{\perp }\to 0$ around $l_2^{\prime }$ and $q$,
we obtain for the gluon bremsstrahlung amplitude

$$
\left. A_{2\rightarrow 2+g}^{a_1a_2a_3a_4a_5}(s,s_1,s_2,t_1,t_2)\right|
_{k_{\bot }\rightarrow 0}=\;g^3\;\;\times
$$
$$
\left\{ l^\mu \left[ 2\left( \frac{p_2}{s_2}-\frac{p_1}{s_1}\right) \left(
l_2^{\,\bot ^{\prime }}q\right) -\left( \frac{p_2}{s_2}-\frac{p_1}{s_1}
\right) \left( l_2^{\,\bot ^{\prime }}k\right) +(B_1+B_2)\left( l_2^{\prime
\,\bot ^{\prime }}q\right) \frac \partial {\partial t}\right] _\mu \right.
\times
$$
$$
\begin{array}{r}
\times \;\;\Gamma (-\alpha (t))\left( \left(
C_{a_1a_2a_3a_4a_5}+C_{a_1a_4a_2a_3a_5}\right) (-\alpha ^{\prime }s)^{\alpha
(t)}\right. + \\
\\
\left. +\left( C_{a_1a_3a_2a_4a_5}+C_{a_1a_4a_3a_2a_5}\right) (\alpha
^{\prime }s)^{\alpha (t)}\right)
\end{array}
$$

\begin{equation}
\label{48}
\begin{array}{l}
+2\alpha ^{\prime }(l_4B_1)\left[ \left( \psi (1)+\ln \frac 1{-\alpha
^{\prime }s_1}\right) \right. \times \\
\\
\times \left( l_2^{\,\bot ^{\prime }}q\right) \left(
C_{a_1a_2a_3a_4a_5}(-\alpha ^{\prime }s)^{\alpha
(t)}+C_{a_1a_3a_2a_4a_5}(\alpha ^{\prime }s)^{\alpha (t)}\right) \Gamma
\left( -\alpha \left( t\right) \right)
\end{array}
\end{equation}
$$
\begin{array}{l}
\qquad \qquad +\;\left( \psi (1)+\ln \frac 1{\alpha ^{\prime }s_1}\right)
\times \\
\\
\times \left. \left( l_2^{\,\bot ^{\prime }}q\right) \left(
C_{a_1a_4a_2a_3a_5}(-\alpha ^{\prime }s)^{\alpha
(t)}+C_{a_1a_4a_3a_2a_5}(\alpha ^{\prime }s)^{\alpha (t)}\right) \Gamma
\left( -\alpha \left( t\right) \right) \right]
\end{array}
$$
$$
\begin{array}{l}
+2\alpha ^{\prime }(lB_2)\left[ \left( \psi (1)+\ln \frac 1{-\alpha ^{\prime
}s_2}\right) \right. \times \\
\\
\times \left( l_2^{\,\bot ^{\prime }}q\right) \left(
C_{a_1a_2a_3a_4a_5}(-\alpha ^{\prime }s)^{\alpha
(t)}+C_{a_1a_4a_3a_2a_5}\left( \alpha ^{\prime }s\right) ^{\alpha
(t)}\right) \Gamma \left( -\alpha \left( t\right) \right)
\end{array}
$$
$$
\begin{array}{l}
\qquad \qquad +\;\left( \psi (1)+\ln \frac 1{\alpha ^{\prime }s_2}\right)
\times \\
\\
\times \left. \left. \left( l_2^{\,\bot ^{\prime }}q\right) \left(
C_{a_1a_4a_2a_3a_5}(-\alpha ^{\prime }s)^{\alpha
(t)}+C_{a_1a_3a_2a_4a_5}(\alpha ^{\prime }s)^{\alpha (t)}\right) \Gamma
\left( -\alpha \left( t\right) \right) \right] \right\}
\end{array}
$$

where $\psi \left( x\right) =\frac {\frac \partial {\partial x}\Gamma \left(
x\right)}{\Gamma \left( x\right)}$ and $B_1,B_2$ are defined by $\left(
\ref {23}\right) $. The first term in $\left( \ref{48}\right) $ is the
non-Abelian generalization of Gribov's pole term of order of $O\left( \frac
1{k_{\bot }}\right) $ and some corrections of order $O\left( k_{\bot
}^0\right) $. The contribution from the color anomalous magnetic moment is
suppressed as $O\left( \frac{k_{\bot }^2}{s_2}\right) $ in this kinematics.
The last terms describe the additional $O\left( k_{\bot}^0\right)$
contribution of the intermediate excited states in the $s_1$ and $s_2$
channels. Eq.$\left( \ref{48}\right) $ can be considered as the dispersion
representation in the invariants $s_i$ with the subtraction at $s_i=0$
\cite{5}. The subtraction constants which provide the agreement with low
energy behavior (where the contribution of the intermediate excited states
dies out as $s_i\approx \omega m_{ch}\to 0$) are equal in all channels. They
are $\psi (1)$.

On the other hand, we could expect the difference in the
values of this constants in $s_1$ and $s_2$ channels because of
the presence of the vector particle in the $s_2$ channel. Due to
the external vector particle which is really the first excited
state of the open string the subtraction point at $s_2=0$ means
the vector state in the channel. It lies above the tachyon point
in the spectrum of states so this additional excited tachyon
contribution like $lB_2\left( \frac 1{\alpha ^{\prime
}s_2-1}+1\right) \left( l_2^{\,\bot ^{\prime }}q\right) \Gamma \left(
-\alpha \left( t\right) \right) \left( C_{a_1a_2a_3a_4a_5}(-\alpha ^{\prime
}s)^{\alpha (t)}+C_{a_1a_4a_3a_2a_5}\left( \alpha ^{\prime }s\right)
^{\alpha (t)}\right) $ could be in the fragmentation regions where $s_2\sim
\frac 1{\alpha ^{\prime }}$ and could change the subtraction constant like
$\psi (1)-1$. But in this case in the multi-Regge limit where
$s_2\rightarrow \infty $ we would have another tensor and color structure
$\sim lB_2\left( l_2^{\,\bot ^{\prime }}q\right) $ than in $\left(
\ref{48}\right) $ and such a simple interpretation of the
difference of the subtraction constants is wrong.

Look at the expression $\left( \ref{48} \right)$ once more. It
is in full agreement with the amplitude factorization. The
difference from the multi-Regge limit of the Low expression in
(\ref{24a}) implies that the tensor structure $\sim -\left(
\frac{p_2}{s_2}-\frac{p_1}{ s_1}\right) \left( l_2^{\,\bot
^{\prime }}k\right) $ is in the first term in $\left(
\ref{48}\right) $ instead of terms $\sim -l_2^{\,\bot ^{\prime }}-\left(
\frac{p_2}{s_2}+\frac{p_1}{s_1}\right) \left( l_2^{\,\bot ^{\prime
}}k\right) $ in (\ref{24a}) plus the contribution of the intermediate
excited states which is the same as for the gluon production amplitude
in the case of the scalar particles collision.

The explanation of the amplitude factorization can be due
to the fact that the contribution of some intermediate excited
states has another tensor and color structure which allows it to
be mixed with the standard Low corrections and to restore the
correct factorization amplitude behavior. In order to justify
this interpretation we should consider the dispersion relations
in detail.  This is possible in the fragmentation region which
is considered in the next section.

\section{Contribution of intermediate excited states in the fragmentation
region.}

In this section we select the region of scalar invariants which is called
the fragmentation region (see Fig.9).
\begin{equation}
\label{49}s_1\sim s\gg m_{ch}^2\quad ,\quad s_2\sim s_{2^{\prime }}\sim
m_{ch}^2
\end{equation}
We will use the string model as well as earlier where $m_{ch}^2=\frac
1{\alpha ^{\prime }}$. The starting point in this section is the
expression in $\left( \ref{42} \right) $. We will follow here the
estimation of integrals in Ref.\cite{5}. In the kinematics $\left(
\ref{49}\right) $ the essential region of integration in $\left(
\ref{42}\right) $ is

\begin{equation}
\label{50}\left.
\begin{array}{c}
1-x\equiv \epsilon \sim \frac 1{\alpha ^{\prime }s}\rightarrow 0 \\
\\
\frac{x-y}{1-y}\equiv z\sim 1\rightarrow 1
\end{array}
\right\} \qquad \Rightarrow \qquad \left\{
\begin{array}{c}
x\approx 1-\epsilon \\
\\
y\approx 1-\frac \epsilon {1-z}
\end{array}
\right.
\end{equation}

and the amplitude $A_{2\rightarrow 1+g+g}(s,s_1,s_2,t_1,t_2)$ is given by

\begin{equation}
\label{51}
\begin{array}{c}
A_{2\rightarrow 1+g+g}(s,s_1,s_2,t_1,t_2)\,\sim \\
\\
\sim \int
\frac{d\epsilon }{\epsilon ^2}\exp \left[ -\alpha ^{\prime }s\epsilon
-\alpha ^{\prime }s_1\epsilon \frac z{1-z}\right] \left| z\right| ^{-\alpha
^{\prime }s_2}\left| \epsilon \frac z{1-z}\right| ^{-\alpha ^{\prime
}t_1}\left| \frac{1-z}z\right| ^{-\alpha ^{\prime }t_2} \\  \\
\times \;l_{2\;\nu }^{\,\bot }\cdot \left( l+2\alpha ^{\prime }\left(
q_2\frac z{\left( 1-z\right) ^2}+q_1\frac 1{1-z}\right) \right) ^\nu \times
\\
\\
\times l_{\;\mu }\cdot \left( p_1\epsilon +p_2\frac{\left( 1-z\right) ^2}
z+q_2\frac{1-z}z\right) ^\mu
\end{array}
\end{equation}
Now, according to the kinematics $\left( \ref{13}\right) $ we can take
\begin{equation}
\label{52}s_1\quad \ll \quad s.
\end{equation}

Then in this asymptotics there are two essential integration regions in
formula in $\left( \ref{51}\right) $
\begin{equation}
\label{53a}1.)\quad \alpha ^{\prime }s_1\epsilon \frac z{z-1}\sim 1
\end{equation}

\begin{equation}
\label{53b}2.)\quad \qquad \frac z{z-1}\sim 1\;.
\end{equation}

The amplitude receives two sorts of contributions which come from regions $
\left( \ref{53a}\right) $ and $\left( \ref{53b}\right) $, respectively.
\begin{equation}
\label{54}A=A_1+A_2
\end{equation}

where $A_1$ and $A_2$ possess different analytic properties. We estimate the
contribution of the regions when $\alpha ^{\prime }s\sim \alpha ^{\prime
}s_1\sim \infty $ and then consider the limit $\frac{s_1}s\rightarrow 0.$

It is worth noting that the Chan-Paton factors appear with their signs which
reflect a number of twists for tree amplitudes \cite{8}. Physical states of
definite mass are eigenstates of the twist operator with eigenvalue $\left(
-1\right) ^N$ where $N=-1$ for the tachyons and $N=0$ for the massless
vector particle.

In the region $\left( \ref{53a}\right) $ we take

\begin{equation}
\label{55}\left.
\begin{array}{c}
\alpha ^{\prime }s\epsilon =x_1 \\
\\
\alpha ^{\prime }s_1\epsilon \frac z{z-1}=x_2
\end{array}
\right\} \qquad \Rightarrow \qquad \left\{
\begin{array}{c}
\epsilon =
\frac{x_1}{\alpha ^{\prime }s} \\  \\
z=\frac 1{\alpha ^{\prime }s_1\left( 1+\frac{x_1}{x_2}\right) }
\end{array}
\right. \quad .
\end{equation}

The contribution $A_1$ is reduced to already well-known analytic and
Chan-Paton structure from the multi-Regge computation (compare with $\left(
\ref{45}\right)$ and see Fig.7 for the Chan-Paton factors)

$$
A_1=\quad g^3\times
$$
$$
\times \;\left\{ C_{a_1a_2a_3a_4a_5}(-\alpha ^{\prime }s)^{\alpha
(t_2)}(-\alpha ^{\prime }s_1)^{\alpha (t_1)-\alpha (t_2)}\right.
$$
\begin{equation}
\label{56}+\;C_{a_1a_4a_2a_3a_5}(-\alpha ^{\prime }s)^{\alpha (t_2)}(\alpha
^{\prime }s_2)^{\alpha (t_1)-\alpha (t_2)}
\end{equation}
$$
+\;\;C_{a_1a_3a_2a_4a_5}(\alpha ^{\prime }s)^{\alpha (t_2)}(-\alpha ^{\prime
}s_1)^{\alpha (t_1)-\alpha (t_2)}
$$
$$
\left. +\;C_{a_1a_4a_3a_2a_5}(\alpha ^{\prime }s)^{\alpha (t_2)}(\alpha
^{\prime }s_1)^{\alpha (t_1)-\alpha (t_2)}\right\} \cdot J_1\left( \frac{s_1}
s,s_2,t_1,t_2\right)
$$

where

\begin{equation}
\label{57}
\begin{array}{c}
J_1\left(
\frac{s_1}s,s_2,t_1,t_2\right) =-\int \int_0^\infty dx_1dx_2x_1^{-2-\alpha
^{\prime }t_2}e^{-x_1}x_2^{-1+\alpha ^{\prime }\left( t_2-t_1\right)
}e^{-x_2}\frac{x_1}{x_2}\left( 1-\frac{s_1x_1}{sx_2}\right) \\  \\
\times \quad l_{2\;\nu }^{\;\bot }\cdot \left( l
\frac{s_1}s+2\alpha ^{\prime }\left( q_2\left( \frac{x_2}{x_1}\right) ^2+q_1
\frac{s_1x_1}{sx_2}\right) \right) ^\nu \\  \\
l_{\;\mu }\cdot \left( \frac{x_1}{\alpha ^{\prime }s_1}p_1+\frac{s_1x_1}{
sx_2 }p_2+\frac{x_1}{x_2}q_2\right) ^\mu
\end{array}
\ .
\end{equation}

{}From the second region in $\left( \ref{53b}\right) $

\begin{equation}
\label{58}\left\{
\begin{array}{c}
\epsilon =\frac 1{\alpha ^{\prime }s}x_1 \\
\\
z=z
\end{array}
\right.
\end{equation}

taking into account the Chan-Paton factors (see Fig.10) we obtain

\begin{equation}
\label{59}
\begin{array}{c}
A_2=g^3\quad \times \\
\\
\times \ \left\{ \left[ C_{a_1a_2a_3a_4a_5}\left( -\alpha ^{\prime }s\right)
^{\alpha \left( t_1\right) }+C_{a_1a_4a_3a_2a_5}\left( \alpha ^{\prime
}s\right) ^{\alpha \left( t_1\right) }\right] \Phi _1\left(
\frac{s_1}s,s_2,t_1,t_2\right) \right. \\  \\
-\left[ C_{a_1a_4a_2a_3a_5}\left( -\alpha ^{\prime }s\right) ^{\alpha \left(
t_1\right) }+C_{a_1a_3a_2a_4a_5}\left( \alpha ^{\prime }s\right) ^{\alpha
\left( t_1\right) }\right] \Phi _2\left(
\frac{s_1}s,s_2,t_1,t_2\right) \\  \\
-\left. \left[ C_{a_1a_2a_4a_3a_5}\left( -\alpha ^{\prime }s\right) ^{\alpha
\left( t_1\right) }+C_{a_1a_3a_4a_2a_5}\left( \alpha ^{\prime }s\right)
^{\alpha \left( t_1\right) }\right] \Phi _3\left( \frac{s_1}
s,s_2,t_1,t_2\right) \right\}
\end{array}
\end{equation}

where

\begin{equation}
\label{60}
\begin{array}{r}
\Phi _1\left(
\frac{s_1}s,s_2,t_1,t_2\right) =-\int_0^\infty dx_1\int_0^1dzx_1^{-2-\alpha
^{\prime }t_1}e^{-x_1-\frac{s_1}sx_1\frac z{1-z}}\times \\  \\
\times z^{-\alpha ^{\prime }s_2}\left( \frac z{1-z}\right) ^{-\alpha
^{\prime }t_1}V_2\left( x,z\right) \\
\\
\Phi _2\left(
\frac{s_1}s,s_2,t_1,t_2\right) =-\int_0^\infty dx_1\int_1^\infty
dzx_1^{-2-\alpha ^{\prime }t_1}e^{-x_1-\frac{s_1}sx_1\frac z{1-z}}\times \\
\\
\times z^{-\alpha ^{\prime }s_2}\left( -\frac z{1-z}\right) ^{-\alpha
^{\prime }t_1}V_2\left( x,z\right) \\
\\
\Phi _3\left(
\frac{s_1}s,s_2,t_1,t_2\right) =-\int_0^\infty dx_1\int_{-\infty
}^0dzx_1^{-2-\alpha ^{\prime }t_1}e^{-x_1-\frac{s_1}sx_1\frac z{1-z}}\times
\\  \\
\times \left( -z\right) ^{-\alpha ^{\prime }s_2}\left( -\frac z{1-z}\right)
^{-\alpha ^{\prime }t_1}V_2\left( x,z\right) \\
\\
V_2\left( x,z\right) =l_{2\;\nu }^{\bot }\left( l+2\alpha ^{\prime }\left(
q_2\frac z{\left( 1-z\right) ^2}+q_1\frac 1{1-z}\right) \right) ^\nu \\
\\
l_{\mu }\left( p_1\frac{x_1}{\alpha ^{\prime }s}+p_2\frac{\left(
1-z\right) ^2}z+q_2\frac{1-z}z\right) ^\mu .
\end{array}
\end{equation}

In the limit $\frac{s_1}s\rightarrow 0$ we have

\begin{equation}
\label{61}
\begin{array}{l}
\left. J_1\left(
\frac{s_1}s,s_2,t_1,t_2\right) \right| _{\frac{s_1}s\rightarrow 0}=2\alpha
^{\prime }\left( l_2^{\bot }q_2\right) \left( l_4B_1\right) \cdot K_0 \\  \\
\left. \Phi _1\left(
\frac{s_1}s,s_2,t_1,t_2\right) \right| _{\frac{s_1}s\rightarrow 0}=\alpha
^{\prime }A_{423}\left( s_2,s_{2^{\prime }},t_1,t_2\right) \cdot K_1 \\  \\
\left. \Phi _2\left(
\frac{s_1}s,s_2,t_1,t_2\right) \right| _{\frac{s_1}s\rightarrow 0}=\alpha
^{\prime }A_{423}\left( s_2,s_{2^{\prime }},t_1,t_2\right) \cdot K_2 \\  \\
\left. \Phi _3\left( \frac{s_1}s(s_2,t_1,t_2\right) \right| _{\frac{s_1}
s\rightarrow 0}=\alpha ^{\prime }A_{423}\left( s_2,s_{2^{\prime
}},t_1,t_2\right) \cdot K_3
\end{array}
\end{equation}

where

\begin{equation}
\label{62}
\begin{array}{l}
K_0=\Gamma \left( -\alpha \left( t_2\right) \right) \Gamma \left( \alpha
\left( t_2\right) -\alpha \left( t_1\right) \right) \\
\\
K_1=-\Gamma \left( -\alpha \left( t_1\right) \right)
\frac{\Gamma \left( 1-\alpha ^{\prime }s_{2^{\prime }}\right) \Gamma \left(
\alpha \left( t_1\right) -\alpha \left( t_2\right) \right) }{\Gamma \left(
2-\alpha ^{\prime }s_2\right) } \\  \\
K_2=-\Gamma \left( -\alpha \left( t_1\right) \right)
\frac{\Gamma \left( -1+\alpha ^{\prime }s_{2^{\prime }}\right) \Gamma \left(
\alpha \left( t_1\right) -\alpha \left( t_2\right) \right) }{\Gamma \left(
\alpha ^{\prime }s_{2^{\prime }}\right) } \\  \\
K_3=\Gamma \left( -\alpha \left( t_1\right) \right) \frac{\Gamma \left(
-1+\alpha ^{\prime }s_2\right) \Gamma \left( 1-\alpha ^{\prime }s_{2^{\prime
}}\right) }{\Gamma \left( 1+\alpha ^{\prime }\left( t_2-t_1\right) \right) }
\end{array}
\end{equation}

and

\begin{equation}
\label{63}
\begin{array}{c}
A_{423}\left( s_2,s_{2^{\prime }},t_1,t_2\right) =\left( t_1-t_2\right)
\left( l_2^{\bot }l\right) +2\left( l_2^{\bot }k\right) \left( lq_2\right) +
\\
\\
+2\alpha ^{\prime }s_2\left( lB_2\right) \left( l_2^{\bot }q_2\right)
-2\left( lB_{2^{\prime }}\right) \left( l_2^{\bot }q_1\right) \left(
1+\alpha ^{\prime }\left( t_1-t_2\right) \right) ,
\end{array}
\end{equation}

the vectors $B_1,B_2$ and $B_{2^{\prime }}$ were given earlier (see
(\ref{23})).  In the asymptotics $\left( 13\right) $ we have
$t_2\rightarrow t_1$ $\left( q_2\rightarrow q_1\right) $ and can take
physically realizable set $\left( \ref{20}\right) $ of on-shell momenta
\{$p_i^{\prime }$\}. In order to keep the gauge invariance we also have to
make the change $l_2\rightarrow l_2^{\prime }$ $\left( l_2^{\prime
}p_2^{\prime }=0\right)$

\begin{equation}
\label{65}
\begin{array}{c}
l_2^{\bot }q_2=l_2^{\bot ^{\prime }}q+
\frac{l_2^{\bot ^{\prime }}k}2 \\  \\
l_2^{\bot }q_1=l_2^{\bot ^{\prime }}q-\frac{l_2^{\bot ^{\prime }}k}2.
\end{array}
\end{equation}

It should be stressed that in the physical region where $s_2\geq 0$ and
$s_{2^{\prime }}\leq 0$ the expression $\left.  \Phi _2\left(
\frac{s_1}s,s_2,t_1,t_2\right) \right| _{\frac{s_1} s\rightarrow 0}$ has
poles only in $s_2$ while $\left. \Phi _1\left( \frac{s_1}
s,s_2,t_1,t_2\right) \right| _{\frac{s_1}s\rightarrow 0}$ has poles only in
$ s_{2^{\prime }}$ (see Eqs.$\left( \ref{61}\right) $ and $\left( \ref{62}
\right) )$. It reflects the fact that these expressions describe the
contribution of the different channels in the amplitude. One of them
describes the contribution of the emission of vector and scalar particles
($ s_{2^{\prime }}$ -channel) and another describes the contribution of the
emission of two vector ($s_2$ -channel).

In this way Eq.$\left( \ref{61}\right) $ can be simplified in accordance
with different regions of $s_i$ (the results of the simplification of the
corresponding functions are represented in the appendix). Substituting Eqs.$
\left(\ref{66}\right) $, $\left( \ref{67}\right) $, $\left( \ref{70}\right) $
and $\left( \ref{73}\right) $ into Eqs.$\left( \ref{57}\right) $, $\left(
\ref{59} \right) $ we obtain for the gluon production amplitude in the
region $\alpha ^{\prime }s_2\sim \alpha ^{\prime }s_{2^{\prime }}\sim 0$ in
the Gribov limit (see $\left( \ref{13}\right) ):$

$$
\left. A\left( s,s_1,s_2,t_1,t_2\right) \right| _{_{\alpha ^{\prime }s_2\sim
0}^{\alpha ^{\prime }s,\ \alpha ^{\prime }s_1\rightarrow \infty }}=
$$
$$
{}
$$
$$
\begin{array}{c}
I \\
\\
=2g^3\left\{ \left[ -2\frac{lp_1}{s_1}l_2^{\bot ^{\prime }}q-\frac
12l_2^{\bot ^{\prime }}l-\frac{lp_1}{s_1}l_2^{\bot ^{\prime }}k+\left(
lB_1\right) \left( l_2^{\bot ^{\prime }}q\right) \frac \partial {\partial
t}\right] \right. \times
\end{array}
$$
$$
{}
$$
$$
\begin{array}{r}
\times \Gamma \left( -\alpha \left( t\right) \right) \left( \left( -\alpha
^{\prime }s\right) ^{\alpha \left( t\right) }\frac 12\left(
C_{a_1a_2a_3a_4a_5}+C_{a_1a_4a_2a_3a_5}\right) \right. + \\
\\
\left. +\left( \alpha ^{\prime }s\right) ^{\alpha \left( t\right) }\frac
12\left( C_{a_1a_3a_2a_4a_5}+C_{a_1a_4a_3a_2a_5}\right) \right) +
\end{array}
$$
$$
II
$$
$$
\begin{array}{c}
+\alpha ^{\prime }\left( lB_1\right) \left[ \left( \psi \left( 1\right)
+\log \frac 1{-\alpha ^{\prime }s_1}\right) \left( l_2^{\bot ^{\prime
}}q\right) \Gamma \left( -\alpha \left( t\right) \right) \times \right.
\qquad \qquad \\
\\
\qquad \qquad \qquad \qquad \times \left( \left( -\alpha ^{\prime }s\right)
^{\alpha \left( t\right) }C_{a_1a_2a_3a_4a_5}+\left( \alpha ^{\prime
}s\right) ^{\alpha \left( t\right) }C_{a_1a_3a_2a_4a_5}\right) + \\
\\
+\left( \psi \left( 1\right) +\log \frac 1{\alpha ^{\prime }s_1}\right)
\left( l_2^{\bot ^{\prime }}q\right) \Gamma \left( -\alpha \left( t\right)
\right) \times \\
\\
\qquad \qquad \qquad \qquad \times \left. \left( \left( -\alpha ^{\prime
}s\right) ^{\alpha \left( t\right) }C_{a_1a_4a_2a_3a_5}+\left( \alpha
^{\prime }s\right) ^{\alpha \left( t\right) }C_{a_1a_4a_3a_2a_5}\right)
\right] +
\end{array}
$$
\begin{equation}
\label{75a}III
\end{equation}
$$
\qquad +\left[ -2\frac{lp_3}{s_{2^{\prime }}}l_2^{\bot ^{\prime }}q-\frac
12l_2^{\bot ^{\prime }}l+\frac{lp_3}{s_{2^{\prime }}}l_2^{\bot ^{\prime
}}k+\left( lB_{2^{\prime }}\right) \left( l_2^{\bot ^{\prime }}q\right)
\frac \partial {\partial t}\right] \times
$$
$$
{}
$$
$$
\begin{array}{r}
\times \Gamma \left( -\alpha \left( t\right) \right) \left( \left( -\alpha
^{\prime }s\right) ^{\alpha \left( t\right) }\frac 12\left(
C_{a_1a_2a_3a_4a_5}+C_{a_1a_2a_4a_3a_5}\right) \right. + \\
\\
\left. +\left( \alpha ^{\prime }s\right) ^{\alpha \left( t\right) }\frac
12\left( C_{a_1a_4a_3a_2a_5}+C_{a_1a_3a_4a_2a_5}\right) \right) +
\end{array}
$$
$$
IV
$$
$$
+\left[ 2\frac{lp_2}{s_2}l_2^{\bot ^{\prime }}q-\frac 12l_2^{\bot ^{\prime
}}l-\frac{lp_2}{s_2}l_2^{\bot ^{\prime }}k-\frac{\left( t_1-t_2\right)
l_2^{\,\bot ^{\prime }}l+2\left( l_2^{\bot ^{\prime }}k\right) lq^{\bot }}{
s_2}+\right.
$$
$$
+\left. \left( lB_2\right) \left( l_2^{\prime \bot ^{\prime }}q\right) \frac
\partial {\partial t}\right] \times
$$
$$
{}
$$
$$
\begin{array}{r}
\times \Gamma \left( -\alpha \left( t\right) \right) \left( \left( -\alpha
^{\prime }s\right) ^{\alpha \left( t\right) }\frac 12\left(
C_{a_1a_4a_2a_3a_5}-C_{a_1a_2a_4a_3a_5}\right) \right. + \\
\\
\left. \left. +\left( \alpha ^{\prime }s\right) ^{\alpha \left( t\right)
}\frac 12\left( C_{a_1a_3a_2a_4a_5}-C_{a_1a_3a_4a_2a_5}\right) \right)
\right\}
\end{array}
$$

Eq.$\left( \ref{75a}\right) $ allows us to single out the contribution of
the nonexcited states in the $s_2$ and $s_{2^{\prime }}$ channel. This
expression can be compared with Eq.$\left( \ref{22}\right) $. In Eq. $\left(
\ref{75a}\right) $ the $I$ and $II$ terms are related to the emission of the
gauge particle from lines 1 and 5, while terms $III$ and $IV$ are associated
with the emission from lines 2 and 3. The $II$ term gives us the
contribution of excited intermediate states in the $s_1$-channel
(it is additional to $ \left( \ref{22}\right) $). Among the
$III$ and $IV$ terms the contributions of such states are absent
and we see that the structure of these terms is in full
agreement with the Low formula. There are only pole leading
terms $\sim O\left( \frac 1{k_{\bot }}\right) $ and
the corrections $\sim O\left( k_{\bot }^0\right) $ coming from
them (compare with terms $IV$ and $VI$ in Eq.(\ref {22})). And
it is worth noting that terms $I$ and $III$ contain the terms $
\sim -\frac 12l_2^{\bot ^{\prime }}l-\frac{lp_i}{s_i}l_2^{\bot
^{\prime }}k$ which are associated with the induced vertex of
interactions of two color scalar particles with two vector
particles. Term $IV$ contains the contribution $\sim -\frac
12l_2^{\bot ^{\prime }}l-\frac{lp_2}{s_2}l_2^{\bot ^{\prime
}}k-\frac{\left( t_1-t_2\right) l_2^{\,\bot ^{\prime }}l+2\left(
l_2^{\bot ^{\prime }}k\right) lq^{\bot }}{s_2}$ which can be
interpreted as the contribution of the color quadrupole electric
moment and of the color magnetic dipole moment (see term $IV$ in
Eq.$\left( \ref{22}\right)$). Thus the amplitude in $\left(
\ref{75a}\right) $ except the contribution of the excited intermediate
states in the $s_1$-channel for bremsstrahlung radiation from scalar
particles (which coincides with \cite{5}) is in full agreement with the
generalization of the Low expressions in $\left( \ref{22}\right) $.

The calculation of the bremsstrahlung amplitude in the fragmentation region $
\alpha ^{\prime }s_2\sim \alpha ^{\prime }s_{2^{\prime }}\sim 1$ in the
Gribov limit can be completed if we substitute the expressions $\left( \ref
{66}\right) ,\left( \ref{68}\right) ,\left( \ref{71}\right) $ and $\left(
\ref{74}\right) $ into Eq.$\left( \ref{57}\right) $ and Eq.$\left( \ref{59}
\right) $.

$$
\left. A\left( s,s_1,s_2,t_1,t_2\right) \right| _{_{\alpha ^{\prime }s_2\sim
\alpha ^{\prime }s_2\sim 1}^{\alpha ^{\prime }s,\ \alpha ^{\prime
}s_1\rightarrow \infty }}=
$$
$$
{}
$$
$$
\begin{array}{c}
I \\
\\
=2g^3\left\{ \left[ -2\frac{lp_1}{s_1}l_2^{\bot ^{\prime }}q-\frac
12l_2^{\bot ^{\prime }}l-\frac{lp_1}{s_1}l_2^{\bot ^{\prime }}k+\left(
lB_1\right) \left( l_2^{\bot ^{\prime }}q\right) \frac \partial {\partial
t}\right] \right. \times
\end{array}
$$
$$
{}
$$
$$
\begin{array}{r}
\times \Gamma \left( -\alpha \left( t\right) \right) \left( \left( -\alpha
^{\prime }s\right) ^{\alpha \left( t\right) }\frac 12\left(
C_{a_1a_2a_3a_4a_5}+C_{a_1a_4a_2a_3a_5}\right) \right. + \\
\\
\left. \left( \alpha ^{\prime }s\right) ^{\alpha \left( t\right) }\frac
12\left( C_{a_1a_3a_2a_4a_5}+C_{a_1a_4a_3a_2a_5}\right) \right) +
\end{array}
$$
$$
II
$$
$$
\begin{array}{c}
+\alpha ^{\prime }\left( lB_1\right) \left[ \left( \psi \left( 1\right)
+\log \frac 1{-\alpha ^{\prime }s_1}\right) \left( l_2^{\bot ^{\prime
}}q\right) \Gamma \left( -\alpha \left( t\right) \right) \times \right.
\qquad \qquad \\
\\
\qquad \qquad \qquad \qquad \times \left( \left( -\alpha ^{\prime }s\right)
^{\alpha \left( t\right) }C_{a_1a_2a_3a_4a_5}+\left( \alpha ^{\prime
}s\right) ^{\alpha \left( t\right) }C_{a_1a_3a_2a_4a_5}\right) + \\
\\
+\left( \psi \left( 1\right) +\log \frac 1{\alpha ^{\prime }s_1}\right)
\left( l_2^{\bot ^{\prime }}q\right) \Gamma \left( -\alpha \left( t\right)
\right) \times \\
\\
\qquad \qquad \qquad \qquad \times \left. \left( \left( -\alpha ^{\prime
}s\right) ^{\alpha \left( t\right) }C_{a_1a_4a_2a_3a_5}+\left( \alpha
^{\prime }s\right) ^{\alpha \left( t\right) }C_{a_1a_4a_3a_2a_5}\right)
\right] +
\end{array}
$$
$$
III
$$
$$
\qquad +\left[ -2\frac{lp_3}{s_{2^{\prime }}}l_2^{\bot ^{\prime }}q-\frac
12l_2^{\bot ^{\prime }}l+\frac{lp_3}{s_{2^{\prime }}}l_2^{\bot ^{\prime
}}k+\left( lB_{2^{\prime }}\right) \left( l_2^{\bot ^{\prime }}q\right)
\frac \partial {\partial t}\right] \times
$$
$$
\times \Gamma \left( -\alpha \left( t\right) \right) \left( \left( -\alpha
^{\prime }s\right) ^{\alpha \left( t\right) }C_{a_1a_2a_3a_4a_5}+\left(
\alpha ^{\prime }s\right) ^{\alpha \left( t\right)
}C_{a_1a_4a_3a_2a_5}\right) +
$$
\begin{equation}
\label{76}IV
\end{equation}
$$
+\left[ 2\left( \frac 1{\alpha ^{\prime }s_{2^{\prime }}-1}+1\right) \left(
\frac 12l_2^{\bot ^{\prime }}l-\frac{lp_3}{s_{2^{\prime }}}l_2^{\bot
^{\prime }}k\right) +\right.
$$
$$
\begin{array}{c}
+\left. 2\alpha ^{\prime }\left( lB_{2^{\prime }}\right) \left( \psi \left(
1\right) -\psi \left( 1-\alpha ^{\prime }s_{2^{\prime }}\right) \right)
\left( l_2^{\bot ^{\prime }}q\right) \right] \Gamma \left( -\alpha \left(
t\right) \right) \times \qquad \qquad \\
\\
\qquad \qquad \qquad \qquad \times \left( \left( -\alpha ^{\prime }s\right)
^{\alpha \left( t\right) }C_{a_1a_2a_3a_4a_5}+\left( \alpha ^{\prime
}s\right) ^{\alpha \left( t\right) }C_{a_1a_4a_3a_2a_5}\right) +
\end{array}
$$
$$
V
$$
$$
+\left[ 2\frac{lp_2}{s_2}l_2^{\bot ^{\prime }}q-\frac 12l_2^{\bot ^{\prime
}}l-\frac{lp_2}{s_2}l_2^{\bot ^{\prime }}k+\left( lB_2\right) \left(
l_2^{\bot ^{\prime }}q\right) \frac \partial {\partial t}\right] \times
$$
$$
{}
$$
$$
\times \Gamma \left( -\alpha \left( t\right) \right) \left( \left( -\alpha
^{\prime }s\right) ^{\alpha \left( t\right) }C_{a_1a_4a_2a_3a_5}+\left(
\alpha ^{\prime }s\right) ^{\alpha \left( t\right)
}C_{a_1a_3a_2a_4a_5}\right) +
$$
$$
VI
$$
$$
+\left[ 2\left( \frac 1{\alpha ^{\prime }s_2-1}+1\right) \left( \frac
12l_2^{\bot ^{\prime }}l+\frac{lp_2}{s_2}l_2^{\bot ^{\prime }}k\right)
+\right.
$$
$$
\begin{array}{c}
+\left. 2\alpha ^{\prime }\left( lB_2\right) \left( \psi \left( 1\right)
-\psi \left( 1+\alpha ^{\prime }s_2\right) \right) \left( l_2^{\bot ^{\prime
}}q\right) \right] \Gamma \left( -\alpha \left( t\right) \right) \times
\qquad \qquad \\
\\
\qquad \qquad \qquad \qquad \times \left( \left( -\alpha ^{\prime }s\right)
^{\alpha \left( t\right) }C_{a_1a_2a_3a_4a_5}+\left( \alpha ^{\prime
}s\right) ^{\alpha \left( t\right) }C_{a_1a_4a_3a_2a_5}\right) -
\end{array}
$$
$$
VII
$$
$$
-2\left( \frac{lp_2}{s_2}+\frac{lp_3}{s_{2^{\prime }}}\right) \left(
l_2^{\bot ^{\prime }}q\right) \frac{\pi \alpha ^{\prime }s_2}{\sin \pi
\alpha ^{\prime }s_2}\times
$$
$$
\times \left. \Gamma \left( -\alpha \left( t\right) \right) \left( \left(
-\alpha ^{\prime }s\right) ^{\alpha \left( t\right)
}C_{a_1a_2a_4a_3a_5}+\left( \alpha ^{\prime }s\right) ^{\alpha \left(
t\right) }C_{a_1a_3a_4a_2a_5}\right) \right\}
$$

Terms $I$ and $II$ describe the emission of gauge particle from the scalar
lines 1 and 5. They coincide with the same terms in the
multi-Regge kinematics since $s_1\gg \frac 1{\alpha ^{\prime }}$ (see
Eq.$\left( \ref{75a}\right) $ ). Term $I$ in Eq.$\left( \ref{76}\right) $
is a non-Abelian generalization of Low's formula \cite{2} (compare with
Eq.$\left( \ref{22}\right) $). The second term in Eq.$\left(
\ref{76}\right) $ represents the additional contribution of excited
intermediate states in the $s_1$ channel. Terms $ III,IV$ and $V,VI$
describe the emission from scalar line 3 and vector line 2 respectively.
Also terms $III$ and $V$ are the non-Abelian generalizations of Low's
formula while the $IV$ and $VI$ ones give us the additional contribution of
the excited intermediate states in the $ s_{2^{\prime }}$ and $s_2$
channels in the fragmentation region where $ \alpha ^{\prime }s_2\sim
\alpha ^{\prime }s_{2^{\prime }}\sim 1$. As far as term $VII$ is concerned
it represents the $u$-channel contribution of the elastic amplitude. To
interpret the results we used that

\begin{equation}
\label{77}\psi \left( 1\right) -\psi \left( 1+\alpha ^{\prime }s_2\right)
=\sum_{n=1}^\infty \left( \frac 1{\alpha ^{\prime }s_2+n}-\frac 1n\right)
\end{equation}
\begin{equation}
\label{78}\psi \left( 1\right) -\psi \left( 1-\alpha ^{\prime }s_{2^{\prime
}}\right) =\sum_{n=1}^\infty \left( \frac 1{-\alpha ^{\prime }s_{2^{\prime
}}+n}-\frac 1n\right)
\end{equation}
and also
\begin{equation}
\label{79}\lim _{\alpha ^{\prime }s_i\rightarrow \infty }\psi \left( 1\pm
\alpha ^{\prime }s_i\right) \rightarrow \log \left( \pm \alpha ^{\prime
}s_i\right) .
\end{equation}

Due to relations (\ref{77}),(\ref{78}) and (\ref{79}) it is
obvious that the inelastic amplitude in the fragmentation region
has infinitely many poles in the $s_{2^{\prime }}$ and $s_2$
channels. In this way formulae (\ref{48}) and (\ref{76})
can be considered as dispersion representations in $s_i$
invariants with subtraction at $s_i=0$. Inspecting (\ref{76})
we note that in general the residues of these poles are equal to
each other. These pieces with the equal residues represent the
contributions of the same intermediate excited states as for the
bremsstrahlung amplitude in the case of the scalar particle
collision (see \cite{5}). But in the $ s_{2^{\prime }}$ channel
(see term $IV$ in Eq.$\left( \ref{76}\right) $) we have the
additional pole contribution at $\alpha ^{\prime }s_{2^{\prime
}}=1$ ($m^2=0$) which is associated with the additional excited
intermediate vector state and the residue in this pole is
different from the contributions of other excited intermediate
states in this channel in comparison with the gluon production amplitude
for the scalar collisions (see \cite{5}). In the $s_2$ channel we
have also the additional pole contribution at $\alpha ^{\prime
}s_2=1$ which can be interpreted as the contribution of the
excited intermediate tachyon state ($m^2=-\frac 1{\alpha
^{\prime }}$) and the residue in this pole is different from the
other one. As seen from Eq.$\left( \ref{76}\right) $ and $\left(
\ref{77}\right) $ the subtraction constant of the dispersion
representation is given by the set of the subtraction constants for each
pole contribution of the intermediate excited states in the corresponding
$s_2$ or $s_{2^{\prime }}$ channel. Its value is determined by two additive
constants $\psi \left( 1\right) $ and $-1$ but with different
tensor factors. In doing so, in the $s_2$ channel the $\psi
\left( 1\right)$ gives us the contribution to the subtraction
constant from those excited states which lie in the mass
spectrum above the external vector state while $-1$ is related to
the excited intermediate tachyon which lies below external
gluon in the mass spectrum. In the $s_{2^{\prime }}$ channel both
constants $\psi \left( 1\right)$ and $-1$ are associated with
the excited intermediate states in the mass spectrum which lie
above the external tachyon.

Finally, from Eq.$\left( \ref{76}\right) $ in the limit $\alpha ^{\prime
}s_2\sim \alpha ^{\prime }s_{2^{\prime }}\sim \infty $ we return to Eq.$
\left( \ref{48}\right) $ for each of the t-channel invariant amplitude. As a
check on this result one can also obtain the same one if one
substitutes expressions $\left( \ref{66}\right) ,\left(
\ref{69}\right) ,\left( \ref{72} \right) $ and $\left(
\ref{75}\right) $ into Eqs.$\left( \ref{57}\right) $ and $\left(
\ref{59}\right) $.

In the limit $s_2\rightarrow \infty $ the contributions of the subtraction
constants of the additional intermediate excited tachyon and vector states
are provided with such a color and tensor structure  which
allows them to be mixed with contributions $\sim $ $\left|
k^{\bot }\right| ^0$ of the nonexcited states. In doing so, we
observe that the terms $\sim l_2^{\bot ^{\prime }}l$ which are
in contradiction with the amplitude factorization are cancelled
by the contributions of the additional intermediate excited
states in the $s_2$ and $s_{2^{\prime }}$ channels, more exactly, due to
their subtraction constants. The correct amplitude
factorization behavior is restored. For the field theory it
means that it is necessary to take into account additionally to
pole diagrams the contributions of the other one in which the
contribution of the scalar and vector intermediate states is in
the corresponding channels. Since in the field theory we take
into account the only pole diagrams it is the reason why in the
multi-Regge kinematics the structure of expression in $ \left(
\ref{24a}\right) $ for the field theory and Eq.$\left( \ref{48}
\right) $ differ from each other.  And we see that
the corresponding subtraction constant for the additional
intermediate vector and scalar states in the $ s_{2^{\prime }}$
and $s_2$ channel is used to restore the correct amplitude
factorization while the subtraction constant of the other
excited intermediate states is kept. In this way the
additive constant really does not depend on the nature of
external particles from which the gluon is radiated.

\section{Summary}

Thus, in this paper we have shown that in the high energy collision of
scalar and vector particles the production amplitude of
the massless gauge particles with the small transverse momentum
does not coincide with the non-Abelian generalization of the Low
formulae. These formulae are in contradiction with the
correct amplitude factorization. The mechanism of the
restoration of the amplitude factorization implies that the
additional intermediate scalar and vector states are excited in
the $s_2$ and $s_{2^{\prime}}$ channels respectively. The other
intermediate states are in full agreement with the same ones as
for the gluon production amplitude for the scalar collision
\cite{5}. In the Regge model these terms in the amplitudes have
non-pole singularities in the invariants $s_i$ (see also
Eq.$\left( \ref{48}\right) $ in the multi-Regge kinematics. The
additive constant (which characterizes the mass spectrum of the
intermediate states, the regime of the Regge behavior and which fits the
amplitude behavior at high energy of the radiated gluon to the amplitude
behavior at low energy of the radiated gluon) does not depend on the
nature of the external particles i.e. whether scalar or vector
particles are the sources of the gluon radiation.  Also we have
demonstrated that the production amplitude of the massless gauge
particles with small transverse momentum in the fragmentation
region of initial particles has nontrivial analytical and other
interesting properties in these invariants $s_i$ (for example,
the equality of the residues of the poles is independent of the
nature of the external particles). The Low-Gribov theorem and its
generalization are used in the field theory to obtain evolution
equations for scattering amplitudes in the Regge kinematics
\cite{10}. In turn, the evolution equations are the starting
point of the effective two-dimensional theory \cite{11}. It can
be equivalent to ordinary QCD.  The explanation of the structure
of terms entering into the amplitude allows us to obtain
important information for building the whole effective field
theory. In particular, from Eq.$ \left( \ref{76}\right) $ we can
obtain vertices like 'reggeon and two or three particles'.

\section{Acknowledgments}

The author is indebted to L.N.Lipatov, A.P.Bukhvostov and
V.A.Kudryavtsev for numerous and stimulating discussions. The
work is supported by grant 93-02-16809 of the Russian Foundation
of Fundamental Research.

\appendix
\section{}
Here we represent the result of calculations of the functions $J_1\left(
\frac{s_1}s,s_2,t_1,t_2\right) $ and $\Phi _i\left( \frac{s_1}
s,s_2,t_1,t_2\right) $ $i=1,2,3$ in the different kinematical regions.

\begin{equation}
\label{66}
\begin{array}{r}
\\
\alpha ^{\prime }s_1\sim \infty \qquad \qquad \qquad \qquad \qquad \qquad
\qquad \\
\\
\left. J_1\right| _{_{\left| k^{\bot }\right| ^2\ll \frac 1{\alpha ^{\prime
}}}^{
\frac{s_1}s\rightarrow \infty }}=\alpha ^{\prime }\Gamma \left( -\alpha
\left( t\right) \right) \left[ -\frac{2\left( lB_1\right) \left( l_2^{\bot
^{\prime }}q\right) }{2\alpha ^{\prime }x}-\frac{\left( lB_1\right) \left(
l_2^{\bot ^{\prime }}k\right) }{2\alpha ^{\prime }x}+\right. \\  \\
\left. +\left( lB_1\right) \left( l_2^{\bot ^{\prime }}q\right) \chi \left(
t\right) \right]
\end{array}
\end{equation}
\begin{equation}
\label{67}
\begin{array}{r}
\\
\alpha ^{\prime }s_{2^{\prime }}\sim 0\qquad \qquad \qquad \qquad \qquad
\qquad \qquad \\
\\
\left. \Phi _1\right| _{_{\left| k^{\bot }\right| ^2\ll \frac 1{\alpha
^{\prime }}}^{
\frac{s_1}s\rightarrow \infty }}=\alpha ^{\prime }\Gamma \left( -\alpha
\left( t\right) \right) \left[ -\frac 1{\alpha ^{\prime }}\left( l_2^{\bot
^{\prime }}l-\frac{lp_3}{s_{2^{\prime }}}\left( l_2^{\bot ^{\prime
}}k\right) \right) -\frac{\left( lq\right) \left( l_2^{\bot ^{\prime
}}k\right) }{2\alpha ^{\prime }x}+\right. \\  \\
\left. +\left( lB_{2^{\prime }}\right) \left( l_2^{\bot ^{\prime }}q\right)
\left( \frac 1{\alpha ^{\prime }x}-\psi \left( \alpha \left( t\right)
\right) \right) \right]
\end{array}
\end{equation}
\begin{equation}
\label{68}
\begin{array}{r}
\\
\alpha ^{\prime }s_{2^{\prime }}\sim 1\qquad \qquad \qquad \qquad \qquad
\qquad \qquad \\
\\
\left. \Phi _1\right| _{_{\left| k^{\bot }\right| ^2\ll \frac 1{\alpha
^{\prime }}}^{
\frac{s_1}s\rightarrow \infty }}=\alpha ^{\prime }\Gamma \left( -\alpha
\left( t\right) \right) \left[ \frac{\left( l_2^{\bot ^{\prime }}l-\frac{
lp_3 }{s_{2^{\prime }}}\left( l_2^{\bot ^{\prime }}k\right) \right) \left(
\alpha ^{\prime }s_{2^{\prime }}+1\right) }{\alpha ^{\prime }\left( \alpha
^{\prime }s_{2^{\prime }}-1\right) }-\right. \qquad \qquad \qquad \\  \\
\left. -\frac{\left( lq\right) \left( l_2^{\bot ^{\prime }}k\right) }{
2\alpha ^{\prime }x}+2\left( lB_{2^{\prime }}\right) \left( l_2^{\bot
^{\prime }}q\right) \left( \frac 1{2\alpha ^{\prime }x}+\chi \left( t\right)
-\psi \left( 1-\alpha ^{\prime }s_{2^{\prime }}\right) \right) \right]
\end{array}
\end{equation}
\begin{equation}
\label{69}
\begin{array}{r}
\\
\alpha ^{\prime }s_{2^{\prime }}\sim \infty ;\qquad s_2\approx s_{2^{\prime
}}\ ,\;B_2\approx B_{2^{\prime }}\qquad \qquad \qquad \\
\\
\left. \Phi _1\right| _{_{\left| k^{\bot }\right| ^2\ll \frac 1{\alpha
^{\prime }}}^{
\frac{s_1}s\rightarrow \infty }}=\alpha ^{\prime }\Gamma \left( -\alpha
\left( t\right) \right) \left[ \frac{lp_2}{\alpha ^{\prime }s_2}l_2^{\bot
^{\prime }}k-\frac{\left( lq\right) \left( l_2^{\bot ^{\prime }}k\right) }{
2\alpha ^{\prime }x}+\right. \\  \\
\left. +2\left( lB_{2^{\prime }}\right) \left( l_2^{\bot ^{\prime }}q\right)
\left( \frac 1{2\alpha ^{\prime }x}+\chi \left( t\right) -\log \left(
-\alpha ^{\prime }s_{2^{\prime }}\right) \right) \right]
\end{array}
\end{equation}
\begin{equation}
\label{70}
\begin{array}{r}
\\
\alpha ^{\prime }s_2\sim 0\qquad \qquad \qquad \qquad \qquad \qquad \qquad
\qquad \\
\\
\left. \Phi _2\right| _{_{\left| k^{\bot }\right| ^2\ll \frac 1{\alpha
^{\prime }}}^{
\frac{s_1}s\rightarrow \infty }}=-\alpha ^{\prime }\Gamma \left( -\alpha
\left( t\right) \right) \left[ -\frac 1{\alpha ^{\prime }}\left( l_2^{\bot
^{\prime }}l-\frac{lp_2}{s_2}\left( l_2^{\bot ^{\prime }}k\right) \right) -
\frac{\left( lq\right) \left( l_2^{\bot ^{\prime }}k\right) }{2\alpha
^{\prime }x}+\right. \\  \\
\left. +2\frac{x\left( l_2^{\bot ^{\prime }}l\right) +\left( lq\right)
\left( l_2^{\bot ^{\prime }}k\right) }{\alpha ^{\prime }s_2}+\left(
lB_2\right) \left( l_2^{\bot ^{\prime }}q\right) \left( \frac 1{\alpha
^{\prime }x}-\psi \left( \alpha \left( t\right) \right) \right) \right]
\end{array}
\end{equation}
\begin{equation}
\label{71}
\begin{array}{r}
\\
\alpha ^{\prime }s_2\sim 1\qquad \qquad \qquad \qquad \qquad \qquad \qquad
\qquad \\
\\
\left. \Phi _2\right| _{_{\left| k^{\bot }\right| ^2\ll \frac 1{\alpha
^{\prime }}}^{
\frac{s_1}s\rightarrow \infty }}=\alpha ^{\prime }\Gamma \left( -\alpha
\left( t\right) \right) \left[ \frac{\left( l_2^{\bot ^{\prime }}l-\frac{
lp_2 }{s_2}\left( l_2^{\bot ^{\prime }}k\right) \right) \left( \alpha
^{\prime }s_2+1\right) }{\alpha ^{\prime }\left( \alpha ^{\prime
}s_2-1\right) } -\right. \qquad \qquad \qquad \\  \\
\left. -\frac{\left( lq\right) \left( l_2^{\bot ^{\prime }}k\right) }{
2\alpha ^{\prime }x}+2\left( lB_2\right) \left( l_2^{\bot ^{\prime
}}q\right) \left( \frac 1{2\alpha ^{\prime }x}+\chi \left( t\right) -\psi
\left( 1+\alpha ^{\prime }s_2\right) \right) \right]
\end{array}
\end{equation}
\begin{equation}
\label{72}
\begin{array}{r}
\\
\alpha ^{\prime }s_2\sim \infty ;\qquad s_2\approx s_{2^{\prime }}\
,\;B_2\approx B_{2^{\prime }}\qquad \qquad \qquad \qquad \\
\\
\left. \Phi _2\right| _{_{\left| k^{\bot }\right| ^2\ll \frac 1{\alpha
^{\prime }}}^{
\frac{s_1}s\rightarrow \infty }}=\alpha ^{\prime }\Gamma \left( -\alpha
\left( t\right) \right) \left[ \frac{lp_2}{\alpha ^{\prime }s_2}l_2^{\bot
^{\prime }}k-\frac{\left( lq\right) \left( l_2^{\bot ^{\prime }}k\right) }{
2\alpha ^{\prime }x}+\right. \\  \\
\left. +2\left( lB_2\right) \left( l_2^{\bot ^{\prime }}q\right) \left(
\frac 1{2\alpha ^{\prime }x}+\chi \left( t\right) -\log \left( \alpha
^{\prime }s_{2^{\prime }}\right) \right) \right]
\end{array}
\end{equation}
\begin{equation}
\label{73}
\begin{array}{l}
\\
\qquad \qquad \qquad \qquad \qquad \qquad \alpha ^{\prime }s_2\sim \alpha
^{\prime }s_{2^{\prime }}\sim 0 \\
\\
\left. \Phi _3\right| _{_{\left| k^{\bot }\right| ^2\ll \frac 1{\alpha
^{\prime }}}^{
\frac{s_1}s\rightarrow \infty }}=\Gamma \left( -\alpha \left( t\right)
\right) \left[ \left( \frac{lp_2}{s_2}+\frac{lp_3}{s_{2^{\prime }}}\right)
\times \right. \\  \\
\left. \times \left( 2l_2^{\prime \bot ^{\prime }}q-l_2^{\prime \bot
^{\prime }}k-2\alpha ^{\prime }x\left( l_2^{\prime \bot ^{\prime }}q\right)
\psi \left( \alpha \left( t\right) \right) \right) -2\frac{x\left( l_2^{\bot
^{\prime }}l\right) +\left( lq\right) \left( l_2^{\bot ^{\prime }}k\right) }{
s_2}\right]
\end{array}
\end{equation}
\begin{equation}
\label{74}
\begin{array}{c}
\\
\alpha ^{\prime }s_2\sim \alpha ^{\prime }s_{2^{\prime }}\sim 1 \\
\\
\left. \Phi _3\right| _{_{\left| k^{\bot }\right| ^2\ll \frac 1{\alpha
^{\prime }}}^{\frac{s_1}s\rightarrow \infty }}=\Gamma \left( -\alpha \left(
t\right) \right) \left( \frac{lp_2}{s_2}+\frac{lp_3}{s_{2^{\prime }}}\right)
\left( l_2^{\bot ^{\prime }}q\right) \frac{2\pi \alpha ^{\prime }s_2}{\sin
\pi \alpha ^{\prime }s_2}
\end{array}
\end{equation}
\begin{equation}
\label{75}
\begin{array}{c}
\\
\alpha ^{\prime }s_2\sim \alpha ^{\prime }s_{2^{\prime }}\sim \infty \\
\\
\left. \Phi _3\right| _{_{\left| k^{\bot }\right| ^2\ll \frac 1{\alpha
^{\prime }}}^{\frac{s_1}s\rightarrow \infty }}=0\quad \left(
go\;to\;\;\infty \;\;along\;real\;axis\;in\;imaginary\;plane\right)
\end{array}
\end{equation}
where
$$
\begin{array}{c}
t_1-t_2\equiv x=-k^{\bot }q \\
\\
s_{2^{\prime }}=s_2+t_1-t_2=s_2+x \\
\\
\chi \left( t\right) =\psi \left( 1\right) -\frac 12\psi \left( -\alpha
\left( t\right) \right) \\
\\
\psi \left( x\right) =\frac{\Gamma ^{\prime }\left( x\right) }{\Gamma \left(
x\right) }\quad \psi \left( 1+x\right) =\psi \left( x\right) +\frac 1x
\end{array}
$$

\input FEYNMAN \bigphotons
\newpage


\begin{center}
\begin{picture}(5000,15000)
\put(1000,7000){\oval(4000,2000)}
\startphantom
\drawline\fermion[\NW\REG](0,7000)[1000]
\stopphantom
\drawline\fermion[\NW\REG](\pbackx,\pbacky)[4000]
\global\Xone=\pbackx     \global\Yone=\pbacky
\global\advance\Xone by -1500
\THICKLINES
\drawarrow[\SE\ATBASE](\pmidx,\pmidy)
\THINLINES
\startphantom
\drawline\fermion[\E\REG](0,7000)[2000]
\stopphantom
\startphantom
\drawline\fermion[\NE\REG](2000,7000)[1000]
\stopphantom
\drawline\fermion[\NE\REG](\pbackx,\pbacky)[4000]
\global\Xtwo=\pbackx     \global\Ytwo=\pbacky
\global\advance \Xtwo by 650
\THICKLINES
\drawarrow[\SW\ATBASE](\pmidx,\pmidy)
\THINLINES
\startphantom
\drawline\fermion[\SE\REG](2000,7000)[1000]
\stopphantom
\drawline\fermion[\SE\REG](\pbackx,\pbacky)[4000]
\global\Xthree=\pbackx     \global\Ythree=\pbacky
\global\advance \Xthree by 650
\THICKLINES
\drawarrow[\NW\ATBASE](\pmidx,\pmidy)
\THINLINES
\startphantom
\drawline\fermion[\SW\REG](0,7000)[1000]
\stopphantom
\drawline\fermion[\SW\REG](\pbackx,\pbacky)[4000]
\global\Xfour=\pbackx     \global\Yfour=\pbacky
\global\advance \Xfour by -1500
\THICKLINES
\drawarrow[\NE\ATBASE](\pmidx,\pmidy)
\THINLINES
\drawline\photon[\N\REG](1000,8000)[4]
\global\Xfive=\pbackx     \global\Yfive=\pbacky
\global\advance \Yfive by 500
\THICKLINES
\drawarrow[\SE\ATBASE](\pmidx,\pmidy)
\THINLINES
\put(\Xone,\Yone){$p_4$}
\put(\Xtwo,\Ytwo){$p_3$}
\put(\Xthree,\Ythree){$p_2$}
\put(\Xfour,\Yfour){$p_1$}
\put(\Xfive,\Yfive){$k$}
\put(-1000,9500){$s_{1^{\prime }}$}
\put(2000,9500){$s_{2^{\prime }}$}
\put(-3000,7000){$t_1$}
\put(5000,7000){$t_2$}
\put(1000,4000){$S$}
\end{picture}
\end{center}

\vskip 0.5 in

\begin{center}
\begin{picture}(20000,2000)
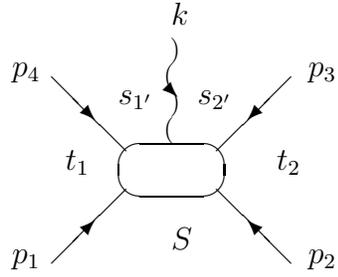
\put(0,1000){Figure 1: The five-particle amplitude.}
\end{picture}
\end{center}

\vskip 0.5 in


\begin{center}
\begin{picture}(5000,14000)
\put(0,7000){\circle{2000}}
\startphantom
\drawline\fermion[\NW\REG](0,7000)[1000]
\stopphantom
\drawline\fermion[\NW\REG](\pbackx,\pbacky)[4000]
\global\Xone=\pbackx     \global\Yone=\pbacky
\global\advance\Xone by -1500
\THICKLINES
\drawarrow[\SE\ATBASE](\pmidx,\pmidy)
\THINLINES
\startphantom
\drawline\fermion[\NE\REG](0,7000)[1000]
\stopphantom
\drawline\fermion[\NE\REG](\pbackx,\pbacky)[4000]
\global\Xtwo=\pbackx     \global\Ytwo=\pbacky
\global\advance \Xtwo by 650
\THICKLINES
\drawarrow[\SW\ATBASE](\pmidx,\pmidy)
\THINLINES
\startphantom
\drawline\fermion[\SE\REG](0,7000)[1000]
\stopphantom
\drawline\fermion[\SE\REG](\pbackx,\pbacky)[4000]
\global\Xthree=\pbackx     \global\Ythree=\pbacky
\global\advance \Xthree by 650
\THICKLINES
\drawarrow[\NW\ATBASE](\pmidx,\pmidy)
\THINLINES
\startphantom
\drawline\fermion[\SW\REG](0,7000)[1000]
\stopphantom
\drawline\fermion[\SW\REG](\pbackx,\pbacky)[4000]
\global\Xfour=\pbackx     \global\Yfour=\pbacky
\global\advance \Xfour by -1500
\THICKLINES
\drawarrow[\NE\ATBASE](\pmidx,\pmidy)
\THINLINES
\put(\Xone,\Yone){$p_4$}
\put(\Xtwo,\Ytwo){$p_3$}
\put(\Xthree,\Ythree){$p_2$}
\put(\Xfour,\Yfour){$p_1$}
\put(0,4000){$s$}
\put(-3000,7000){$t$}
\end{picture}
\end{center}

\vskip 0.5 in

\begin{center}
\begin{picture}(20000,2000)
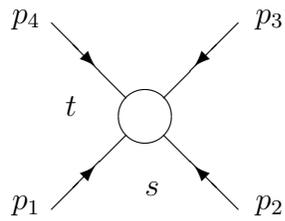
\put(0,1000){Figure 2: The elastic scalar particle amplitude.}
\end{picture}
\end{center}

\newpage


\begin{center}
\global\newsavebox{\ELASTIC} \savebox{\ELASTIC}(0,0) {\
\begin{picture}(5000,10000)
\put(0,5000){\circle{2000}}
\startphantom
\drawline\fermion[\NW\REG](0,5000)[1000]
\stopphantom
\drawline\fermion[\NW\REG](\pbackx,\pbacky)[3000]
\global\Xone=\pbackx     \global\Yone=\pbacky
\global\Xfive=\pmidx     \global\Yfive=\pmidy
\global\advance\Xone by -1200
\startphantom
\drawline\fermion[\NE\REG](0,5000)[1000]
\stopphantom
\drawline\fermion[\NE\REG](\pbackx,\pbacky)[3000]
\global\Xtwo=\pbackx     \global\Ytwo=\pbacky
\global\Xsix=\pmidx     \global\Ysix=\pmidy
\global\advance \Xtwo by 650
\startphantom
\drawline\fermion[\SE\REG](0,5000)[1000]
\stopphantom
\drawline\photon[\SE\REG](\pbackx,\pbacky)[4]
\global\Xthree=\pbackx     \global\Ythree=\pbacky
\global\Xseven=\pmidx     \global\Yseven=\pmidy
\global\advance \Xthree by 650
\startphantom
\drawline\fermion[\SW\REG](0,5000)[1000]
\stopphantom
\drawline\fermion[\SW\REG](\pbackx,\pbacky)[3000]
\global\Xfour=\pbackx     \global\Yfour=\pbacky
\global\Xeight=\pmidx     \global\Yeight=\pmidy
\global\advance \Xfour by -1200
\put(\Xone,\Yone){$p_4$}
\put(\Xtwo,\Ytwo){$p_3$}
\put(\Xthree,\Ythree){$p_2$}
\put(\Xfour,\Yfour){$p_1$}
\end{picture}
}
\begin{picture}(5000,12000)
\drawoldpic\ELASTIC(2500,5000)
\drawline\photon[\N\REG](0,6000)[3]
\THICKLINES
\drawarrow[\NE\ATBASE](\pbackx,\pbacky)
\THINLINES
\global\advance \pbackx by 100  \global\advance \pbacky by 450
\put(\pbackx,\pbacky){$k$}
\put(0,5000){\circle*{2000}}
\put(-1000,1000){$A_{\gamma}$}
\end{picture}
\hskip 0.2in
\begin{picture}(1000,12000)(0,0)
\put(-2000,5000){$=$}
\end{picture}
\hskip 0.2in
\begin{picture}(5000,12000)
\drawoldpic\ELASTIC(2500,5000)
\drawline\photon[\N\REG](0,6000)[3]
\THICKLINES
\drawarrow[\NE\ATBASE](\pbackx,\pbacky)
\THINLINES
\global\advance \pbackx by 100  \global\advance \pbacky by 450
\put(\pbackx,\pbacky){$k$}
\put(-1000,1000){$A^{int}_{\gamma}$}
\end{picture}
\hskip 0.2in
\begin{picture}(1000,12000)(0,0)
\put(-2000,5000){$+$}
\end{picture}
\vskip 0.1in
\begin{picture}(5000,10000)
\drawoldpic\ELASTIC(2500,5000)
\drawline\photon[\NE\REG](\Xfive,\Yfive)[3]
\THICKLINES
\drawarrow[\E\ATBASE](\pbackx,\pbacky)
\THINLINES
\global\advance \pbackx by 450  \global\advance \pbacky by 450
\put(\pbackx,\pbacky){$k$}
\end{picture}
\hskip 0.2in
\begin{picture}(1000,5000)(0,0)
\put(-2000,5000){$+$}
\end{picture}
\hskip 0.2in
\begin{picture}(5000,10000)(0,0)
\drawoldpic\ELASTIC(2500,5000)
\drawline\photon[\NW\REG](\Xsix,\Ysix)[3]
\THICKLINES
\drawarrow[\N\ATBASE](\pbackx,\pbacky)
\THINLINES
\global\advance \pbackx by -450  \global\advance \pbacky by 650
\put(\pbackx,\pbacky){$k$}
\put(5000,0){$A^{ext}_{\gamma}$}
\end{picture}
\hskip 0.2in
\begin{picture}(1000,10000)(0,0)
\put(-2000,5000){$+$}
\end{picture}
\hskip 0.2in
\begin{picture}(5000,10000)(0,0)
\drawoldpic\ELASTIC(2500,5000)
\drawline\photon[\NE\REG](\Xseven,\Yseven)[3]
\THICKLINES
\drawarrow[\E\ATBASE](\pbackx,\pbacky)
\THINLINES
\global\advance \pbackx by 450  \global\advance \pbacky by 450
\put(\pbackx,\pbacky){$k$}
\end{picture}
\hskip 0.2in
\begin{picture}(1000,10000)(0,0)
\put(-1500,5000){$+$}
\end{picture}
\hskip 0.4in
\begin{picture}(5000,10000)(0,0)
\drawoldpic\ELASTIC(2500,5000)
\drawline\photon[\NW\REG](\Xeight,\Yeight)[3]
\THICKLINES
\drawarrow[\N\ATBASE](\pbackx,\pbacky)
\THINLINES
\global\advance \pbackx by 100  \global\advance \pbacky by 650
\put(\pbackx,\pbacky){$k$}
\end{picture}
\end{center}

\vskip 0.5 in
\begin{picture}(20000,2000)
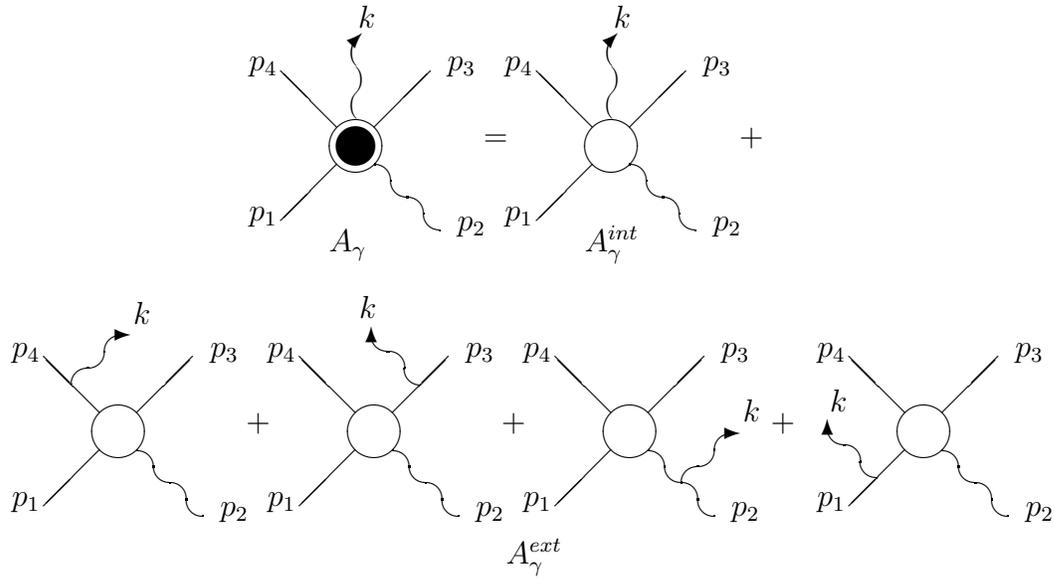
\put(0,1000){Figure 3: The amplitude can be divided into an external and
internal part.}
\end{picture}


\begin{center}
\begin{picture}(5000,14000)
\put(0,7000){\circle{2000}}
\startphantom
\drawline\fermion[\NW\REG](0,7000)[1000]
\stopphantom
\drawline\fermion[\NW\REG](\pbackx,\pbacky)[4000]
\global\Xone=\pbackx     \global\Yone=\pbacky
\global\advance\Xone by -1500
\THICKLINES
\drawarrow[\SE\ATBASE](\pmidx,\pmidy)
\THINLINES
\startphantom
\drawline\fermion[\NE\REG](0,7000)[1000]
\stopphantom
\drawline\fermion[\NE\REG](\pbackx,\pbacky)[4000]
\global\Xtwo=\pbackx     \global\Ytwo=\pbacky
\global\advance \Xtwo by 650
\THICKLINES
\drawarrow[\SW\ATBASE](\pmidx,\pmidy)
\THINLINES
\startphantom
\drawline\fermion[\SE\REG](0,7000)[1000]
\stopphantom
\drawline\gluon[\SE\REG](\pbackx,\pbacky)[2]
\global\Xthree=\pbackx     \global\Ythree=\pbacky
\global\advance \Xthree by 650
\global\advance \pmidx by 250   \global\advance \pmidy by 250
\THICKLINES
\drawarrow[\NW\ATBASE](\pmidx,\pmidy)
\THINLINES
\startphantom
\drawline\fermion[\SW\REG](0,7000)[1000]
\stopphantom
\drawline\fermion[\SW\REG](\pbackx,\pbacky)[4000]
\global\Xfour=\pbackx     \global\Yfour=\pbacky
\global\advance \Xfour by -1500
\THICKLINES
\drawarrow[\NE\ATBASE](\pmidx,\pmidy)
\THINLINES
\put(\Xone,\Yone){$p_4$}
\put(\Xtwo,\Ytwo){$p_3$}
\put(\Xthree,\Ythree){$p_2$}
\put(\Xfour,\Yfour){$p_1$}
\put(0,4000){$s$}
\put(-3000,7000){$t$}
\end{picture}
\end{center}

\begin{picture}(20000,2000)
\put(-2000,1000){Figure 4: The basic elastic amplitude with one vector and
three scalar particles.}
\end{picture}

\newpage


\begin{center}
\begin{picture}(5000,14000)
\drawline\photon[\E\REG](0,7000)[5]
\global\advance\pmidy by 250
\THICKLINES
\drawarrow[\E\ATBASE](\pmidx,\pmidy)
\THINLINES
\global\advance\pmidy by 850
\put(\pmidx,\pmidy){$q$}
\drawline\fermion[\NW\REG](\photonfrontx,\photonfronty)[4000]
\THICKLINES
\drawarrow[\SE\ATBASE](\pmidx,\pmidy)
\THINLINES
\global\advance\fermionbackx by -1200
\put(\fermionbackx,\fermionbacky){$p_4$}
\drawline\fermion[\SW\REG](\photonfrontx,\photonfronty)[4000]
\THICKLINES
\drawarrow[\NE\ATBASE](\pmidx,\pmidy)
\THINLINES
\global\advance\fermionbackx by -1200
\put(\fermionbackx,\fermionbacky){$p_1$}
\drawline\fermion[\NE\REG](\photonbackx,\photonbacky)[4000]
\THICKLINES
\drawarrow[\SW\ATBASE](\pmidx,\pmidy)
\THINLINES
\global\advance\fermionbackx by 650
\put(\fermionbackx,\fermionbacky){$p_3$}
\drawline\gluon[\SE\FLIPPED](\photonbackx,\photonbacky)[2]
\global\advance\pmidx by -250   \global\advance\pmidy by -250
\THICKLINES
\drawarrow[\NW\ATBASE](\pmidx,\pmidy)
\THINLINES
\global\advance\pbackx by 650
\put(\pbackx,\pbacky){$p_2$}
\put(2500,4000){$s$}
\put(-3000,7000){$t$}
\end{picture}
\end{center}

\vskip 0.5 in
\begin{picture}(20000,2000)
\put(0,1000){Figure 5: The basic elastic process of scattering in the
Regge kinematics.}
\end{picture}


\begin{center}
\begin{picture}(5000,14000)
\put(1000,7000){\oval(4000,2000)}
\startphantom
\drawline\fermion[\NW\REG](0,7000)[1000]
\stopphantom
\drawline\fermion[\NW\REG](\pbackx,\pbacky)[4000]
\global\Xone=\pbackx     \global\Yone=\pbacky
\global\advance\Xone by -1500
\THICKLINES
\drawarrow[\SE\ATBASE](\pmidx,\pmidy)
\THINLINES
\startphantom
\drawline\fermion[\E\REG](0,7000)[2000]
\stopphantom
\startphantom
\drawline\fermion[\NE\REG](2000,7000)[1000]
\stopphantom
\drawline\fermion[\NE\REG](\pbackx,\pbacky)[4000]
\global\Xtwo=\pbackx     \global\Ytwo=\pbacky
\global\advance \Xtwo by 650
\THICKLINES
\drawarrow[\SW\ATBASE](\pmidx,\pmidy)
\THINLINES
\startphantom
\drawline\fermion[\SE\REG](2000,7000)[1000]
\stopphantom
\drawline\gluon[\SE\FLIPPED](\pbackx,\pbacky)[2]
\global\advance\pmidx by -250
\global\advance\pmidy by -250
\global\Xthree=\pbackx     \global\Ythree=\pbacky
\global\advance \Xthree by 650
\THICKLINES
\drawarrow[\NW\ATBASE](\pmidx,\pmidy)
\THINLINES
\startphantom
\drawline\fermion[\SW\REG](0,7000)[1000]
\stopphantom
\drawline\fermion[\SW\REG](\pbackx,\pbacky)[4000]
\global\Xfour=\pbackx     \global\Yfour=\pbacky
\global\advance \Xfour by -1500
\THICKLINES
\drawarrow[\NE\ATBASE](\pmidx,\pmidy)
\THINLINES
\drawline\gluon[\N\REG](1000,8000)[3]
\global\advance\pmidy by 500
\global\Xfive=\pbackx     \global\Yfive=\pbacky
\global\advance \Yfive by 500
\THICKLINES
\drawarrow[\S\ATBASE](\pmidx,\pmidy)
\THINLINES
\put(\Xone,\Yone){$p_5$}
\put(\Xtwo,\Ytwo){$p_3$}
\put(\Xthree,\Ythree){$p_2$}
\put(\Xfour,\Yfour){$p_1$}
\put(\Xfive,\Yfive){$p_4$}
\put(-1000,9500){$s_1$}
\put(2000,9500){$s_2$}
\put(-3000,7000){$t_1$}
\put(5000,7000){$t_2$}
\put(1000,4000){$S$}
\end{picture}
\end{center}

\vskip 0.5 in

\begin{center}
\begin{picture}(20000,2000)
\put(0,1000){Figure 6: The inelastic amplitude.}
\end{picture}
\end{center}

\newpage


\begin{center}
\begin{picture}(10000,14000)
\THICKLINES
\drawline\photon[\E\REG](0,7000)[5]
\global\advance\pmidy by 300
\drawarrow[\E\ATBASE](\pmidx,\pmidy)
\global\advance\pmidy by 1250
\put(\pmidx,\pmidy){$q_1$}
\THINLINES
\drawline\gluon[\N\FLIPPEDCENTRAL](\pbackx,\pbacky)[4]
\global\advance\pbacky by 200
\put(\pbackx,\pbacky){$p_4$}
\put(\pfrontx,\pfronty){\circle*{500}}
\global\advance\pmidx by 550
\THICKLINES
\drawarrow[\S\ATBASE](\pmidx,\pmidy)
\drawline\photon[\E\FLIPPED](\pfrontx,\pfronty)[5]
\global\advance\pmidy by -200
\drawarrow[\E\ATBASE](\pmidx,\pmidy)
\global\advance\pmidy by 300
\global\advance\pmidy by 1250
\put(\pmidx,\pmidy){$q_2$}
\put(\pmidx,10000){$s_2$}
\THINLINES
\drawline\fermion[\NE\REG](\pbackx,\pbacky)[6000]
\THICKLINES
\drawarrow[\SW\ATBASE](\pmidx,\pmidy)
\THINLINES
\global\Xtwo=\pbackx     \global\Ytwo=\pbacky
\global\advance \Xtwo by 200
\drawline\gluon[\SE\FLIPPED](\pfrontx,\pfronty)[4]
\put(\pmidx,7000){$t_2$}
\global\advance\pmidx by -250  \global\advance\pmidy by -250
\THICKLINES
\drawarrow[\NW\ATBASE](\pmidx,\pmidy)
\THINLINES
\global\Xthree=\pbackx     \global\Ythree=\pbacky
\global\advance \Xthree by 200
\drawline\fermion[\NW\REG](0,7000)[6000]
\THICKLINES
\drawarrow[\SE\ATBASE](\pmidx,\pmidy)
\THINLINES
\global\Xone=\pbackx     \global\Yone=\pbacky
\global\advance\Xone by -1200
\drawline\fermion[\SW\REG](\pfrontx,\pfronty)[6000]
\put(-3000,7000){$t_1$}
\THICKLINES
\drawarrow[\NE\ATBASE](\pmidx,\pmidy)
\THINLINES
\global\Xfour=\pbackx     \global\Yfour=\pbacky
\global\advance \Xfour by -1200
\put(\Xone,\Yone){$p_5$}
\put(\Xtwo,\Ytwo){$p_3$}
\put(\Xthree,\Ythree){$p_2$}
\put(\Xfour,\Yfour){$p_1$}
\put(4000,3500){$s$}
\put(2000,10000){$s_1$}
\end{picture}
\end{center}

\vskip 0.5 in
\begin{picture}(20000,2000)
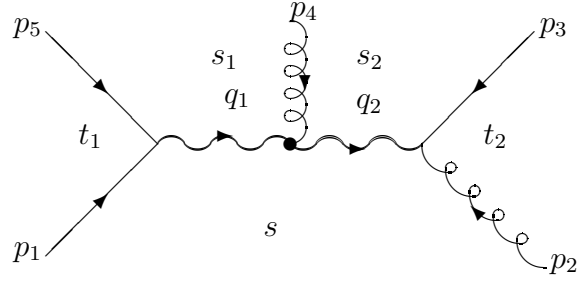
\put(0,1000){Figure 7: The gluon production amplitude in the multi-Regge
kinematics.}
\end{picture}


\begin{center}
\global\newsavebox{\CHAN} \savebox{\CHAN}(0,0) {\
\begin{picture}(0,0)
\drawline\fermion[\S\REG](0,12000)[4000]
\drawline\fermion[\SW\REG](\pbackx,\pbacky)[3000]
\global\advance\pmidx by 500   \global\advance\pmidy by -100
\put(\pmidx,\pmidy){$a_1$}
\drawline\fermion[\E\REG](\pfrontx,\pfronty)[4000]
\bigphotons
\drawline\photon[\SE\REG](\pbackx,\pbacky)[4]
\global\advance \pmidx by -1500   \global\advance \pmidy by -100
\put(\pmidx,\pmidy){$a_2$}
\drawline\fermion[\N\REG](\pfrontx,\pfronty)[4000]
\drawline\fermion[\NE\REG](\pbackx,\pbacky)[3000]
\global\advance \pmidx by -1500   \global\advance \pmidy by -100
\put(\pmidx,\pmidy){$a_3$}
\drawline\fermion[\W\REG](\pfrontx,\pfronty)[4000]
\drawline\fermion[\NW\REG](\pbackx,\pbacky)[3000]
\global\advance \pmidx by 500    \global\advance \pmidy by -100
\put(\pmidx,\pmidy){$a_5$}
\end{picture}
} \global\newsavebox{\PATON} \savebox{\PATON}(0,0) {\
\begin{picture}(0,0)
\drawline\fermion[\S\REG](0,12000)[4000]
\drawline\fermion[\SW\REG](\pbackx,\pbacky)[3000]
\global\advance\pmidx by 500   \global\advance\pmidy by -100
\put(\pmidx,\pmidy){$a_1$}
\global\advance \fermionlength by 1000
\multroothalf\fermionlength
\drawline\fermion[\NE\REG](\pfrontx,\pfronty)[\fermionlength]
\drawline\fermion[\NE\REG](\pbackx,\pbacky)[\fermionlength]
\drawline\fermion[\NE\REG](\pbackx,\pbacky)[3000]
\global\advance \pmidx by -1500   \global\advance \pmidy by -100
\put(\pmidx,\pmidy){$a_3$}
\drawline\fermion[\S\REG](\pfrontx,\pfronty)[4000]
\drawline\photon[\SE\REG](\pbackx,\pbacky)[4]
\global\advance \pmidx by -1500   \global\advance \pmidy by -100
\put(\pmidx,\pmidy){$a_2$}
\multroothalf\fermionlength
\drawline\fermion[\NW\REG](\pfrontx,\pfronty)[\fermionlength]
\drawline\fermion[\NW\REG](\pbackx,\pbacky)[\fermionlength]
\drawline\fermion[\NW\REG](\pbackx,\pbacky)[3000]
\global\advance \pmidx by 500    \global\advance \pmidy by -100
\put(\pmidx,\pmidy){$a_5$}
\end{picture}
}
\begin{picture}(6000,16000)
\drawoldpic\CHAN(0,0)
\bigphotons
\drawline\photon[\N\REG](2000,12000)[3]
\global\advance\pfrontx by 400   \global\advance\pfronty by 500
\put(\pfrontx,\pfronty){$a_4$}
\put(0,3000){$C_{a_1 a_2 a_3 a_4 a_5}$}
\THICKLINES
\drawarrow[\S\ATBASE](0,10000)
\drawarrow[\E\ATBASE](2000,8000)
\drawarrow[\N\ATBASE](4000,10000)
\drawarrow[\W\ATBASE](3000,12000)
\drawarrow[\W\ATBASE](1000,12000)
\THINLINES
\end{picture}
\hskip 0.1 in
\begin{picture}(0,16000)
\put(0,10000){$\bigoplus$}
\end{picture}
\hskip 0.4 in
\begin{picture}(6000,16000)
\drawoldpic\CHAN(0,0)
\bigphotons
\drawline\photon[\N\REG](2000,8000)[6]
\global\advance\pfrontx by -400
\global\advance\pfronty by -1000
\put(\pfrontx,\pfronty){$a_4$}
\put(0,3000){$C_{a_1 a_4 a_2 a_3 a_5}$}
\THICKLINES
\drawarrow[\S\ATBASE](0,10000)
\drawarrow[\E\ATBASE](1000,8000)
\drawarrow[\E\ATBASE](3000,8000)
\drawarrow[\N\ATBASE](4000,10000)
\drawarrow[\W\ATBASE](3000,12000)
\drawarrow[\W\ATBASE](1000,12000)
\THINLINES
\end{picture}
\hskip 0.1 in
\begin{picture}(0,16000)
\put(0,10000){$\bigoplus$}
\end{picture}
\hskip 0.4 in
\begin{picture}(6000,16000)
\drawoldpic\PATON(0,0)
\bigphotons
\drawline\photon[\N\REG](1200,10800)[3]
\global\advance\pfrontx by 400   \global\advance\pfronty by 1000
\put(\pfrontx,\pfronty){$a_4$}
\put(0,3000){$C_{a_1 a_3 a_2 a_4 a_5}$}
\THICKLINES
\drawarrow[\S\ATBASE](0,10000)
\drawarrow[\NE\ATBASE](1000,9000)
\drawarrow[\NE\ATBASE](3000,11000)
\drawarrow[\S\ATBASE](4000,10000)
\drawarrow[\NW\ATBASE](3000,9000)
\drawarrow[\NW\ATBASE](1500,10500)
\drawarrow[\NW\ATBASE](500,11500)
\THINLINES
\end{picture}
\hskip 0.1 in
\begin{picture}(0,16000)
\put(0,10000){$\bigoplus$}
\end{picture}
\hskip 0.4 in
\begin{picture}(6000,16000)
\drawoldpic\PATON(0,0)
\bigphotons
\drawline\photon[\N\REG](2800,10800)[3]
\global\advance\pfrontx by -1200   \global\advance\pfronty by 1200
\put(\pfrontx,\pfronty){$a_4$}
\put(0,3000){$C_{a_1 a_4 a_3 a_2 a_5}$}
\THICKLINES
\drawarrow[\S\ATBASE](0,10000)
\drawarrow[\NE\ATBASE](1000,9000)
\drawarrow[\NE\ATBASE](2500,10500)
\drawarrow[\NE\ATBASE](3500,11500)
\drawarrow[\S\ATBASE](4000,10000)
\drawarrow[\NW\ATBASE](3000,9000)
\drawarrow[\NW\ATBASE](1000,11000)
\THINLINES
\end{picture}
\end{center}

\vskip 0.5 in
\begin{picture}(20000,2000)
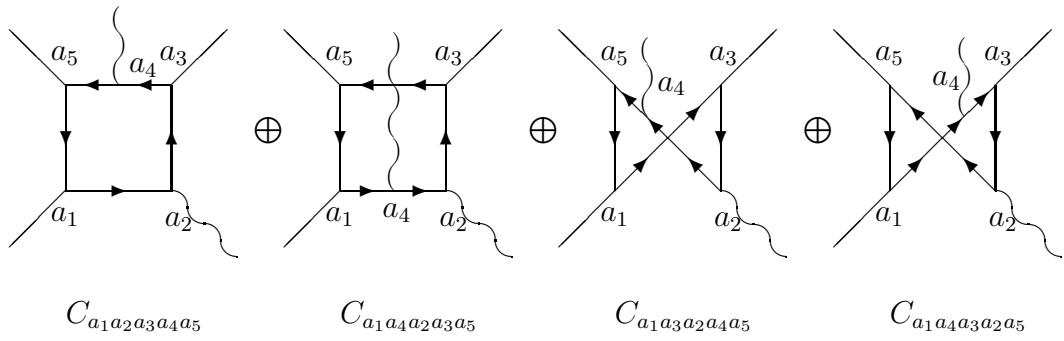
\put(-2000,1000){Figure 8: Chan-Paton factors which contribute in the
multi-Regge kinematics.}
\end{picture}

\newpage


\begin{center}
\begin{picture}(10000,14000)
\THICKLINES
\drawline\photon[\E\REG](0,7000)[6]
\global\advance\pmidx by -600
\global\advance\pmidy by 300
\drawarrow[\E\ATBASE](\pmidx,\pmidy)
\global\advance\pmidy by 600
\put(\pmidx,\pmidy){$q_1$}
\THINLINES
\drawline\gluon[\N\CENTRAL](\pbackx,\pbacky)[4]
\global\advance\pbacky by 200
\put(\pbackx,\pbacky){$p_4$}
\put(\pfrontx,\pfronty){\circle*{500}}
\global\advance\pmidx by -550
\THICKLINES
\drawarrow[\S\ATBASE](\pmidx,\pmidy)
\THINLINES
\drawline\fermion[\NE\REG](\pfrontx,\pfronty)[6000]
\THICKLINES
\drawarrow[\SW\ATBASE](\pmidx,\pmidy)
\THINLINES
\global\Xtwo=\pbackx     \global\Ytwo=\pbacky
\global\advance \Xtwo by 200
\drawline\gluon[\SE\FLIPPED](\pfrontx,\pfronty)[4]
\put(\pmidx,7000){$t_2$}
\global\advance\pmidx by -250  \global\advance\pmidy by -250
\THICKLINES
\drawarrow[\NW\ATBASE](\pmidx,\pmidy)
\THINLINES
\global\Xthree=\pbackx     \global\Ythree=\pbacky
\global\advance \Xthree by 200
\drawline\fermion[\NW\REG](0,7000)[6000]
\THICKLINES
\drawarrow[\SE\ATBASE](\pmidx,\pmidy)
\THINLINES
\global\Xone=\pbackx     \global\Yone=\pbacky
\global\advance\Xone by -1200
\drawline\fermion[\SW\REG](\pfrontx,\pfronty)[6000]
\put(-3000,7000){$t_1$}
\THICKLINES
\drawarrow[\NE\ATBASE](\pmidx,\pmidy)
\THINLINES
\global\Xfour=\pbackx     \global\Yfour=\pbacky
\global\advance \Xfour by -1200
\put(\Xone,\Yone){$p_5$}
\put(\Xtwo,\Ytwo){$p_3$}
\put(\Xthree,\Ythree){$p_2$}
\put(\Xfour,\Yfour){$p_1$}
\put(3000,3500){$s$}
\put(2000,10000){$s_1$}
\put(6700,10000){$s_{2'}$}
\end{picture}
\end{center}

\vskip 0.0 in
\begin{picture}(20000,2000)
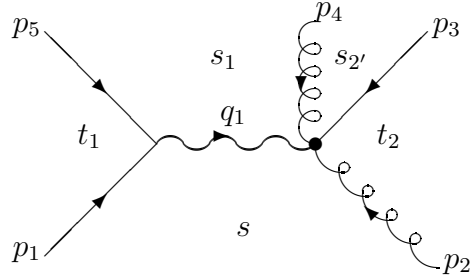
\put(0,2000){Figure 9: Massless particle production amplitude in the
fragmentation}
\put(0,500){region.}
\end{picture}


\begin{center}
\global\newsavebox{\CHAM} \savebox{\CHAM}(0,0) {\
\begin{picture}(0,0)
\drawline\fermion[\S\REG](0,12000)[4000]
\drawline\fermion[\SW\REG](\pbackx,\pbacky)[3000]
\global\advance\pmidx by 500   \global\advance\pmidy by -100
\put(\pmidx,\pmidy){$a_1$}
\drawline\fermion[\E\REG](\pfrontx,\pfronty)[4000]
\bigphotons
\drawline\photon[\SE\REG](\pbackx,\pbacky)[4]
\global\advance \pmidx by -1500   \global\advance \pmidy by -100
\put(\pmidx,\pmidy){$a_2$}
\drawline\fermion[\N\REG](\pfrontx,\pfronty)[4000]
\drawline\fermion[\NE\REG](\pbackx,\pbacky)[3000]
\global\advance \pmidx by -1500   \global\advance \pmidy by -100
\put(\pmidx,\pmidy){$a_3$}
\drawline\fermion[\W\REG](\pfrontx,\pfronty)[4000]
\drawline\fermion[\NW\REG](\pbackx,\pbacky)[3000]
\global\advance \pmidx by 500    \global\advance \pmidy by -100
\put(\pmidx,\pmidy){$a_5$}
\end{picture}
} \global\newsavebox{\PATOM} \savebox{\PATOM}(0,0) {\
\begin{picture}(0,0)
\drawline\fermion[\S\REG](0,12000)[4000]
\drawline\fermion[\SW\REG](\pbackx,\pbacky)[3000]
\global\advance\pmidx by 500   \global\advance\pmidy by -100
\put(\pmidx,\pmidy){$a_1$}
\global\advance \fermionlength by 1000
\multroothalf\fermionlength
\drawline\fermion[\NE\REG](\pfrontx,\pfronty)[\fermionlength]
\drawline\fermion[\NE\REG](\pbackx,\pbacky)[\fermionlength]
\drawline\fermion[\NE\REG](\pbackx,\pbacky)[3000]
\global\advance \pmidx by -1500   \global\advance \pmidy by -100
\put(\pmidx,\pmidy){$a_3$}
\drawline\fermion[\S\REG](\pfrontx,\pfronty)[4000]
\drawline\photon[\SE\REG](\pbackx,\pbacky)[4]
\global\advance \pmidx by -1500   \global\advance \pmidy by -100
\put(\pmidx,\pmidy){$a_2$}
\multroothalf\fermionlength
\drawline\fermion[\NW\REG](\pfrontx,\pfronty)[\fermionlength]
\drawline\fermion[\NW\REG](\pbackx,\pbacky)[\fermionlength]
\drawline\fermion[\NW\REG](\pbackx,\pbacky)[3000]
\global\advance \pmidx by 500    \global\advance \pmidy by -100
\put(\pmidx,\pmidy){$a_5$}
\end{picture}
}
\begin{picture}(6000,14000)
\drawoldpic\CHAM(0,0)
\bigphotons
\drawline\photon[\N\REG](2000,12000)[3]
\global\advance\pfrontx by 400   \global\advance\pfronty by 500
\put(\pfrontx,\pfronty){$a_4$}
\put(0,3000){$C_{a_1 a_2 a_3 a_4 a_5}$}
\THICKLINES
\drawarrow[\S\ATBASE](0,10000)
\drawarrow[\E\ATBASE](2000,8000)
\drawarrow[\N\ATBASE](4000,10000)
\drawarrow[\W\ATBASE](3000,12000)
\drawarrow[\W\ATBASE](1000,12000)
\THINLINES
\end{picture}
\hskip 0.1 in
\begin{picture}(0,14000)
\put(0,10000){$\bigoplus$}
\end{picture}
\hskip 0.4 in
\begin{picture}(6000,14000)
\drawoldpic\PATOM(0,0)
\bigphotons
\drawline\photon[\N\REG](2800,10800)[3]
\global\advance\pfrontx by -1200   \global\advance\pfronty by 1200
\put(\pfrontx,\pfronty){$a_4$}
\put(0,3000){$C_{a_1 a_4 a_3 a_2 a_5}$}
\THICKLINES
\drawarrow[\S\ATBASE](0,10000)
\drawarrow[\NE\ATBASE](1000,9000)
\drawarrow[\NE\ATBASE](2500,10500)
\drawarrow[\NE\ATBASE](3500,11500)
\drawarrow[\S\ATBASE](4000,10000)
\drawarrow[\NW\ATBASE](3000,9000)
\drawarrow[\NW\ATBASE](1000,11000)
\THINLINES
\end{picture}
\hskip 0.1 in
\begin{picture}(0,14000)
\put(0,10000){$\bigoplus$}
\end{picture}
\hskip 0.4 in
\begin{picture}(6000,14000)
\drawoldpic\CHAM(0,0)
\bigphotons
\drawline\photon[\N\REG](2000,8000)[6]
\global\advance\pfrontx by -400
\global\advance\pfronty by -1000
\put(\pfrontx,\pfronty){$a_4$}
\put(0,3000){$C_{a_1 a_4 a_2 a_3 a_5}$}
\THICKLINES
\drawarrow[\S\ATBASE](0,10000)
\drawarrow[\E\ATBASE](1000,8000)
\drawarrow[\E\ATBASE](3000,8000)
\drawarrow[\N\ATBASE](4000,10000)
\drawarrow[\W\ATBASE](3000,12000)
\drawarrow[\W\ATBASE](1000,12000)
\THINLINES
\end{picture}
\hskip 0.1 in
\begin{picture}(0,14000)
\put(0,10000){$\bigoplus$}
\end{picture}
\hskip 0.4 in
\begin{picture}(6000,14000)
\drawoldpic\PATOM(0,0)
\bigphotons
\drawline\photon[\N\REG](1200,10800)[3]
\global\advance\pfrontx by 400   \global\advance\pfronty by 1000
\put(\pfrontx,\pfronty){$a_4$}
\put(0,3000){$C_{a_1 a_3 a_2 a_4 a_5}$}
\THICKLINES
\drawarrow[\S\ATBASE](0,10000)
\drawarrow[\NE\ATBASE](1000,9000)
\drawarrow[\NE\ATBASE](3000,11000)
\drawarrow[\S\ATBASE](4000,10000)
\drawarrow[\NW\ATBASE](3000,9000)
\drawarrow[\NW\ATBASE](1500,10500)
\drawarrow[\NW\ATBASE](500,11500)
\THINLINES
\end{picture}
\hskip 0.1 in
\begin{picture}(0,14000)
\put(0,10000){$\bigoplus$}
\end{picture}
\vskip 0 in
\begin{picture}(0,14000)
\put(0,10000){$\bigoplus$}
\end{picture}
\hskip 0.4 in
\begin{picture}(6000,14000)
\drawoldpic\CHAM(0,0)
\bigphotons
\drawline\photon[\NW\REG](4000,10000)[6]
\global\advance\pfrontx by 400
\put(\pfrontx,\pfronty){$a_4$}
\put(0,3000){$C_{a_1 a_2 a_4 a_3 a_5}$}
\THICKLINES
\drawarrow[\S\ATBASE](0,10000)
\drawarrow[\E\ATBASE](2000,8000)
\drawarrow[\N\ATBASE](4000,11000)
\drawarrow[\N\ATBASE](4000,9000)
\drawarrow[\W\ATBASE](3000,12000)
\drawarrow[\W\ATBASE](1000,12000)
\THINLINES
\end{picture}
\hskip 0.1 in
\begin{picture}(0,14000)
\put(0,10000){$\bigoplus$}
\end{picture}
\hskip 0.4 in
\begin{picture}(6000,14000)
\drawoldpic\PATOM(0,0)
\bigphotons
\drawline\photon[\NW\REG](4000,10000)[6]
\global\advance\pfrontx by 400
\put(\pfrontx,\pfronty){$a_4$}
\put(0,3000){$C_{a_1 a_3 a_4 a_2 a_5}$}
\THICKLINES
\drawarrow[\S\ATBASE](0,10000)
\drawarrow[\NE\ATBASE](1000,9000)
\drawarrow[\NE\ATBASE](2500,10500)
\drawarrow[\NE\ATBASE](3500,11500)
\drawarrow[\S\ATBASE](4000,11000)
\drawarrow[\S\ATBASE](4000,9000)
\drawarrow[\NW\ATBASE](3000,9000)
\drawarrow[\NW\ATBASE](1000,11000)
\THINLINES
\end{picture}
\end{center}

\vskip 0.0 in
\begin{picture}(20000,2000)
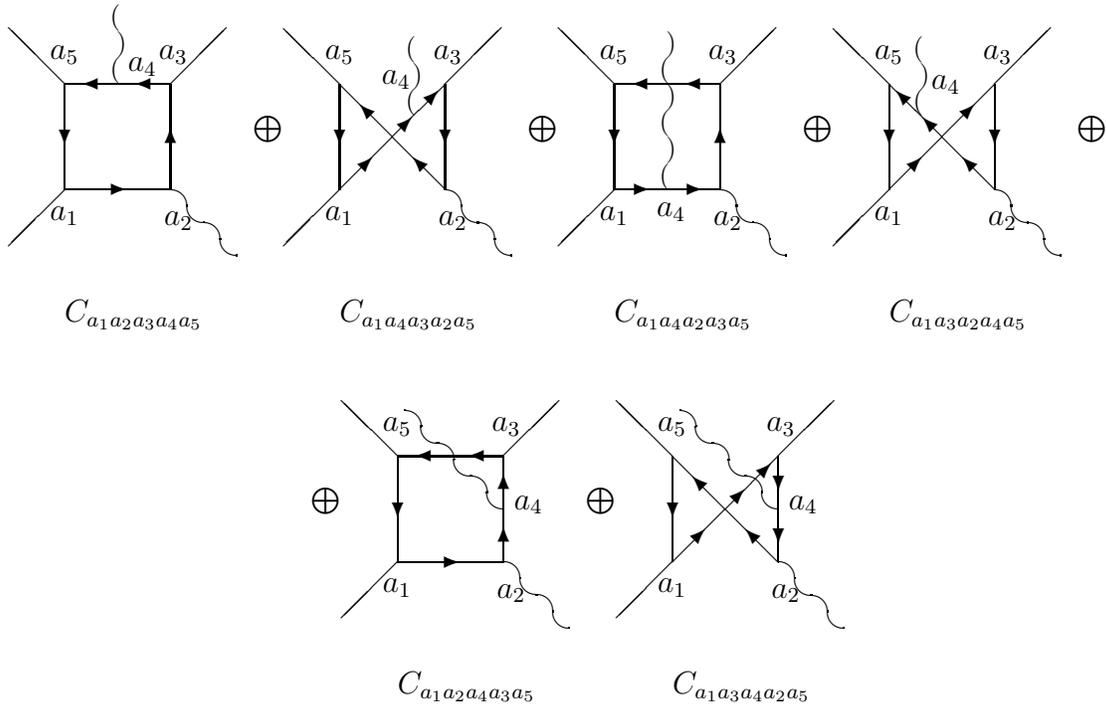
\put(0,2000){Figure 10: Chan-Paton factors which contribute into the $A_2$
piece of}
\put(0,500){ the gluon production amplitude in the fragmentation region.}
\end{picture}

\end{document}